\documentclass[journal]{IEEEtran}

\usepackage{cite}
\usepackage[pdftex]{graphicx}
\graphicspath{{assets/}} 
\DeclareGraphicsExtensions{.pdf,.jpeg,.png}
\usepackage{amsmath, amssymb, amsfonts}
\usepackage{algorithm}
\usepackage{algpseudocode}
\usepackage{authblk}
\usepackage{array}
\usepackage{url}
\usepackage{verbatim}
\usepackage{balance}
\usepackage{subfig}
\usepackage{caption}
\newcommand{\sectionEND}[1]{}

\newlength{\commentlen}

\begin{document}











\title{NeuRehab: A Reinforcement Learning and Spiking Neural Network-Based Rehab Automation Framework}

\author{
    \IEEEauthorblockN{Phani Pavan Kambhampati\IEEEauthorrefmark{1},  Chainesh Gautam\IEEEauthorrefmark{2}, Jagan P\IEEEauthorrefmark{3}, Madhav Rao\IEEEauthorrefmark{4}}
    
    \IEEEauthorblockA{International Institute of Information Technology Bangalore, India}
    
    Email: \{\IEEEauthorrefmark{1}phanipavan.k, \IEEEauthorrefmark{2}chainesh.gautam \IEEEauthorrefmark{3}jagan.p \IEEEauthorrefmark{4}mr\}@iiitb.ac.in
}


\maketitle

\begin{abstract}
    Recent advancements in robotic rehabilitation therapy have provided modular exercise systems for post-stroke muscle recovery with basic control schemes. But these systems struggle to adapt to patients' complex and ever-changing behaviour, and to operate within mobile settings, such as heat and power. 
    To aid this, we present NeuRehab: an end-to-end framework consisting of a training and inference pipeline with AI-based automation, co-designed with neuromorphic computing-based control systems that balance action performance, power consumption, and observed latency. The framework consists of 2 partitions. One is designated for the rehabilitation device based on ultra-low power spiking networks deployed on dedicated neuromorphic hardware. The other resides on stationary hardware that can accommodate computationally intensive hardware for fine-tuning on a per-patient basis.
    By maintaining a communication channel between both the modules and splitting the algorithm components, the power and latency requirements of the movable system have been optimised, while retaining the learning performance advantages of compute- and power-hungry hardware on the stationary machine. 
    As part of the framework, we propose (a) the split machine learning processes for efficiency in architectural utilisation, and (b) task-specific temporal optimisations to lower edge-inference control latency.
    This paper evaluates the proposed methods on a reference stepper motor-based shoulder exercise. Overall, these methods offer comparable performance uplifts over the State-of-the-art for neuromorphic deployment, while achieving over 60\% savings in both power and latency during inference compared to standard implementations.
\end{abstract}


\section{Introduction}
Stroke and other neurological conditions often leave patients with long-term motor impairments that require intensive and repetitive rehabilitation to regain function. Conventional therapy depends on one-to-one supervision by a trained physiotherapist, which is costly, difficult to scale, and hard to sustain over long periods of time. Robotic rehabilitation systems offer programmable, high-intensity therapy, but most existing devices still follow fixed protocols with limited personalisation and little adaptation to the patient’s changing ability.

For shoulder and upper-limb rehabilitation, the control problem is particularly challenging. The device must guide the limb through flexion–extension exercises safely, assist when the patient is weak, yield when the patient moves actively, and avoid excessive forces that might cause pain or discomfort. These requirements are difficult to satisfy with purely hand-tuned controllers or fixed trajectory generators. Reinforcement learning (RL) provides an alternative by optimising a policy directly from interactions, once a reward is defined that encodes clinical goals. In principle, this allows assist-as-needed behaviour that modulates support based on patient effort, joint state, and safety constraints.

Directly training and deploying deep RL on a wheelchair-mounted exoskeleton creates two main obstacles. Training policies on real hardware is unsafe and data inefficient because exploration inevitably produces suboptimal, potentially harmful actions, and the number of safe rollouts is limited. At the same time, standard deep RL implementations assume a powerful processor and a generous power budget, while a wheelchair platform must run on a compact battery with strict thermal and energy limits. Simply placing a GPU near the motors is not practical.

That is where the field of neuromorphic computing fills the gap with a power-efficient compute unit capable of machine learning inference. Specifically, a neuromorphic chip running a spiking neural network (SNN) replaces the traditional artificial neural network (ANN) and a GPU~\cite{neurApps-edgeApps-schuman}. Not only does this consume low power, but SNNs also promise lower latency. Since the wheelchair already houses high-power circuits and motors, using a neuromorphic chip reduces heat generation during computation and adds only a small burden to the existing power budget. NeuroCARE~\cite{neurocare-fwsota-tian} proposed an open-loop deployment pipeline; we improve upon this to provide a closed-loop learning workflow for robotic rehabilitation systems.

In this work, we propose \textit{NeuRehab}: a neuromorphic reinforcement learning framework designed for wheelchair-based
rehabilitation. We build on the XoRehab platform~\cite{xorehab1-rehabintro-jaganp, xorehab2-rehabhand-jaganp}, an IoT-enabled exoskeletal system for motor-sense recovery exercises, and introduce a hardware and software architecture that explicitly separates low-power on-chair inference from compute-heavy training and monitoring on a docking station. This docking station includes powerful GPUs to fine-tune the behavioural model using data acquired from the wheelchair's edge device.

To enable the safe development and analysis of control policies before deployment, we design two custom simulation environments for the shoulder joint. The kinematic environment (KENV) captures the delay-controlled behaviour of stepper motors and their discrete motion profiles. The dynamic environment (DENV) models torque-driven pendulum-like physics with patient interaction. Both follow the Gymnasium API, expose clinically meaningful observations such as joint angles, velocities, patient torque and strain, and use reward terms that penalise excessive force, abrupt motion, and misalignment between the patient and the device.

On the algorithmic side, we use Soft Actor–Critic (SAC) as the base RL method and introduce a heterogeneous training scheme called Hybrid-SAC (HSAC). In HSAC, the policy (actor) is implemented as an SNN, and the critic remains an ANN. This keeps value estimation in a high-precision domain, which stabilises learning, while aligning the actor architecture with the neuromorphic hardware used at inference time. We then propose two inference-time optimisations for spiking control policies. Spiking Post-Training Temporal Quantisation (SPTTQ) treats the number of spiking time steps as a post-training quantisation parameter. The Sequent Leaky (SLeaky) neuron retains its membrane potential across RL steps, reducing the charge-up overhead at each step.

Our main contributions are:

\begin{itemize}
    \item We present a complete hardware and software framework for autonomous rehabilitation on the XoRehab platform. This includes a split wheelchair–dock architecture and two task-specific simulation environments, KENV and DENV, which model stepper-based and torque-based shoulder control.
    \item We propose a heterogeneous neuromorphic RL algorithm, HSAC, that combines a spiking actor with an ANN critic. Across standard MuJoCo tasks and the proposed rehabilitation environments, HSAC tracks the performance of an all-ANN SAC (ASAC) baseline and consistently outperforms a fully spiking SAC (SSAC).
    \item We introduce two inference optimisations for spiking actors in continuous control: SPTTQ and the Sequent Leaky neuron. Together, they reduce spike counts and effective latency while preserving control performance.
    \item We provide an empirical study of the learned policies and spiking dynamics, including analyses of stable and unstable temporal behaviour, spike distributions, and strain and torque profiles in KENV and DENV. This illustrates the proposed methods' trade-off between performance, power, and responsiveness.
\end{itemize}

These components together demonstrate that neuromorphic RL is made feasible for edge-constrained rehabilitation devices when the environment, the learning algorithm, and the hardware are co-designed. The following sections describe the framework architecture, simulation environment design, spiking network methods, and experimental results in detail.

\sectionEND{Introduction}

\section{The Framework}

\begin{figure}[ht!]
    \centering
    \includegraphics[width=1\linewidth]{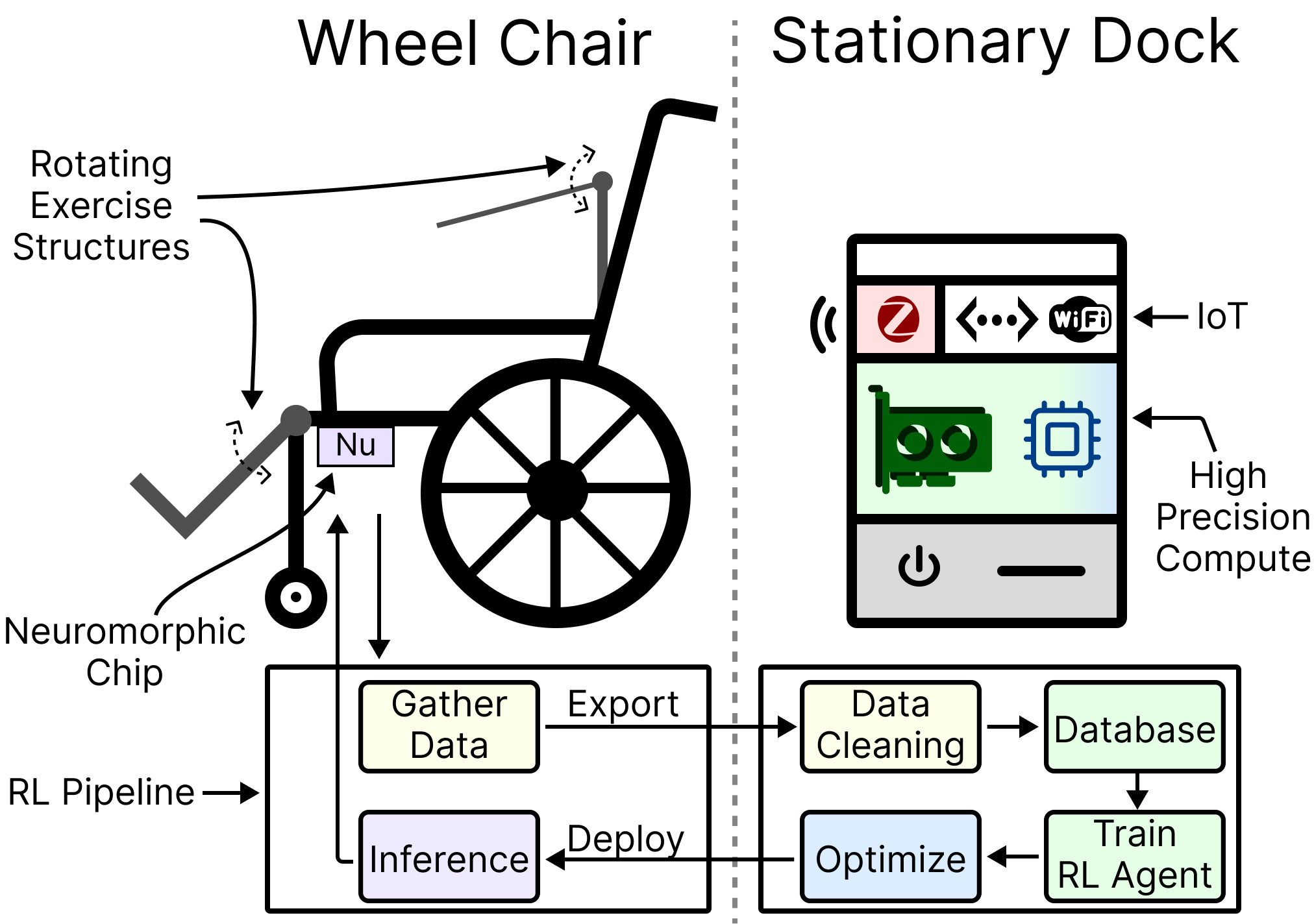}
    \caption{A framework diagram representing the proposed neuromorphic automation system.}
    \label{fig:framework}
\end{figure}

Components in Figure~\ref{fig:framework} form the backbone of an autonomous, modular, self-learning, and configurable theoretical framework for rehabilitation.  
The physical aspect consists of two key components: 
\begin{itemize}
\item The wheelchair: an edge device consisting of control electronics, a neuromorphic compute system, and sensors, and 
\item The docking station: a stationary device that accepts the wheelchair for charging, data acquisition, and model updates. 
\end{itemize}
The software components include an RTOS environment for real-time hardware control, a reinforcement-learning-based automation system, and a cloud-connected software stack for remote management.
This multipart system enables integration across heterogeneous hardware architectures to achieve overall efficiency.

The wheelchair consists of the main rehabilitation hardware, a power supply system, and an ARM-based edge controller board with a neuromorphic coprocessor. The neuromorphic chip performs inference, while the main microcontroller handles the communication and actuations. This model leverages the neuromorphic chip's power efficiency and the widespread support for the ARM microcontroller to run the RTOS control processes.

On the other hand, the docking station houses power-hungry components like a GPU and a general-purpose processor. This allows the dock to maintain a wireless connection to the wheelchair, fine-tune control parameters on the fly, relay data to remote locations, and retrain the model based on patient interactions. This paper focuses on optimising the reinforcement learning component to leverage available resources.

\sectionEND{The Framework}

\section{Hardware Platform}

XoRehab~\cite{xorehab1-rehabintro-jaganp} is chosen as the target platform to implement and test the proposed pipelining architecture. XoRehab is an IoT-enabled rehabilitation system targeting motor-sense recovery exercises for post-stroke patients. These exercises are part of a therapy that includes flexions (bending of a joint) and extensions (straightening of a joint). This rehab system was designed as an exoskeletal add-on to a medical wheelchair and consists of multiple disjoint robotised components. Each component handles the exercise for a part of the patient's body, including wrists, elbows, shoulders, hips, and other joints in the legs.
All the rotating actuations are achieved using either the servo motors 
or stepper motors.

\begin{figure}[ht]
    \centering
    \includegraphics[width=1\linewidth,trim={200 80 300 60},clip]{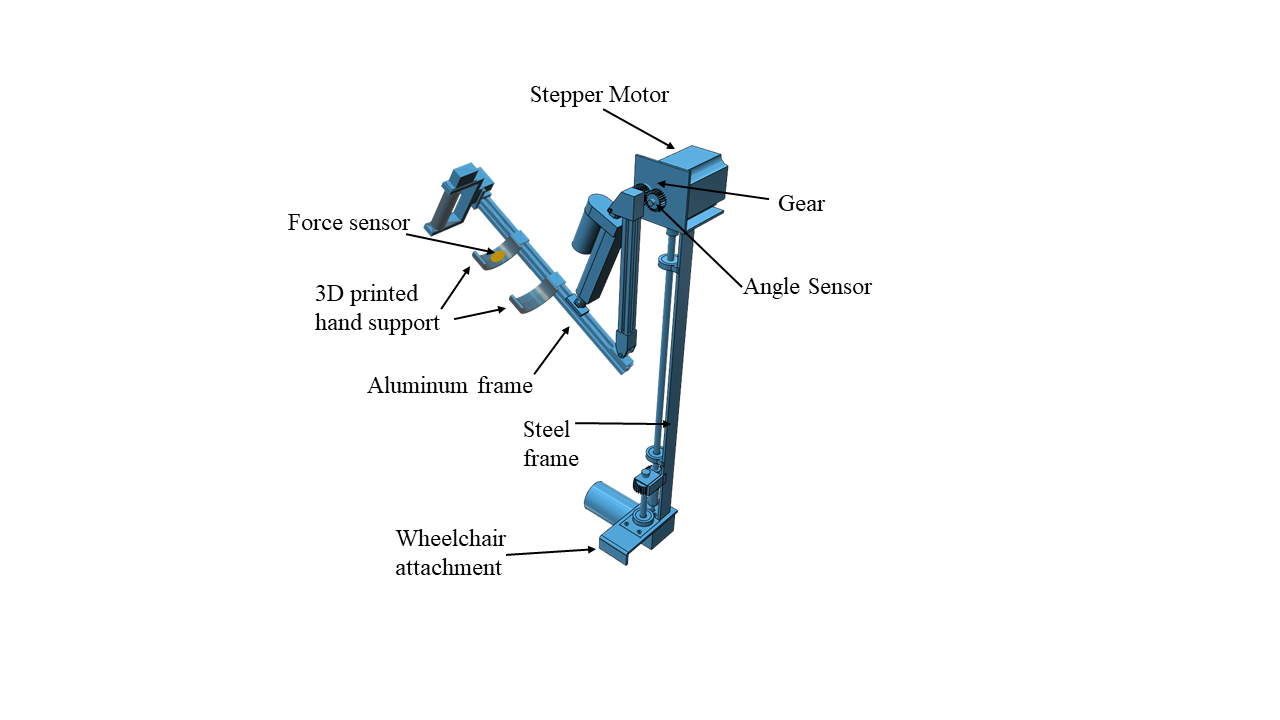}
    \caption{The 3D design of XoRehab upper limb mechanism with force sensor.}
    \label{fig:3D}
\end{figure}

As a proof of concept, this paper validates the framework for flexion and extension exercises of the shoulder joint, which can be extended to other wheelchair components. The exercise involves raising and lowering the hand forward up to a certain angle. A 3D representation of the shoulder attachment is shown in Figure~\ref{fig:3D}. Rather than a DC motor, this joint operates using a stepper motor with a 1:15 gearbox. A standard DC motor operates at a specific torque and at various speeds, with torque and speed controlled by input current and voltage. But a stepper motor's range of operation is discretised into a fixed number of steps and operates using a stepper controller.
This means rotation happens in multiples of fixed step angles. The stepper motor and its controller use target angle and speed as control variables rather than power/torque. A stepper motor provides position-hold functionality, which prevents the arm structure from collapsing under gravity in the absence of a control signal. With additional gear reductions and the inherent construction of a stepper motor, the system requires no power to maintain a particular state. While this is favourable for performing exercises, it also introduces complexity to its software simulation, as discussed later in the paper.

The rehabilitation system also incorporates a robust IoT architecture communicating instantaneous sensor readings from the edge device to local and cloud-based data storage.
The improvements~\cite{xorehab2-rehabhand-jaganp} add machine-learning-based systemic anomaly detection using sensor values. The framework we propose adds to existing capabilities by substituting hard-coded exercise routines with reinforcement learning (RL)- based routines, enabling the system to tune its response to user interaction. The addition of a complex action-pattern learner addresses multiple objectives that a patient desires to perform.

The main goal for the current RL system is to aid in the following tasks:

\begin{itemize}
	\item Help the patient to move the hand up/down.
	\item Not exert too much force on the hand if the patient struggles to perform the exercise.
	\item Modulate the speed based on the patient's preference in real time.
	\item Follow the hand/not resist if the patient is performing the exercise faster.
\end{itemize}

Real-time patient feedback is essential for automating this procedure. While bio-signals such as electromyogram (EMG) can be used to monitor strain feedback, a stroke often impairs muscle responses. Reading EMG in such a scenario will result in noisy signals. Hence, feedback from the patient must be acquired from external sources. To aid this, the system is modified to include a couple of force sensors in the target shoulder component. Output from these sensors serves as the primary feedback from the patient, reflecting the system-imposed movement on the patient's arm and the patient's ability to complete the action. 
And to automate the exercise process, we use a reinforcement learning architecture to train a control scheme, as described below.

\sectionEND{Hardware Platform}

\section{Simulating in Software}

Reinforcement learning (RL) \cite{schulman2017proximalpolicyoptimizationalgorithms, 712192} is a framework in which an agent learns to make decisions by interacting with an environment and receiving feedback in the form of rewards. Unlike proportional - integral - derivative (PID) controllers, which follow fixed rules tuned by hand, RL adapts its behaviour through experience. This adaptability makes RL more suitable for rehabilitation settings, where patient responses are uncertain, and system dynamics may change over time. Deep RL extends these ideas by using neural networks to approximate value functions and policies \cite{schulman2017proximalpolicyoptimizationalgorithms, haarnoja2018softactorcriticoffpolicymaximum}, enabling the agent to learn effective strategies even in high-dimensional and continuous control problems.

Training RL agents directly on hardware is unsafe because learning requires heavy exploration, including many unstable or suboptimal actions. A simulator provides a safe and repeatable environment in which the agent can explore widely, learn from mistakes, and refine its policy without risking patients or hardware. With these motivations in place, we now describe the two simulation environments developed for our study.

\subsection{Simulation Environments}

We designed two custom environments that model the arm at different levels of abstraction, namely the Kinematic environment and the Dynamic environment. Both are implemented using the Gymnasium API \cite{towers2025gymnasiumstandardinterfacereinforcement}, making them compatible with standard reinforcement learning libraries. Each defines an action space, an observation space, and standard methods for resetting and stepping through the simulation. Rendering is also supported to visualise the arm’s motion.

\subsubsection{Kinematic Environment}
The kinematic environment abstracts away the underlying physics and focuses instead on the arm's higher-level movement. The agent’s action is a delay value in milliseconds, which controls the stepping rate of a stepper motor. 
The delay is proportional to the stepper's angular velocity. The observation space includes normalised values such as applied forces, angular position, angular velocity, and patient torque. Patient torque can be constant or sinusoidal, while the system follows an assist-as-needed rule that supplies motor torque only when the patient’s contribution is insufficient. The reward combines speed-profile adherence, blockage penalties, over-forcing penalties, smoothness terms, and patient-effort rewards. Episodes terminate when the motor completes its allotted number of steps.

This environment reflects the discrete, quantised behaviour of stepper motors. Since the arm structure moves in fixed increments of angle, fine-grained continuous control is not directly possible. This is similar to integer quantisation in machine learning, where continuous optimisation becomes difficult in discrete spaces. Smoothness must arise from the temporal pattern of delays rather than from continuous torque outputs.

\subsubsection{Dynamic Environment}
The dynamic environment models the arm structure as a torque-controlled physical system, similar to a DC or servo motor-based actuation stage. The agent directly outputs a torque command, and the simulator combines this with gravitational, inertial, frictional, and patient-applied torques. These forces are integrated using pendulum-like model equations to compute angular acceleration, velocity, and position. A parallel hand-only trajectory is simulated to measure strain, representing the deviation between system-driven and patient-driven motion. The observation includes angle, velocity, patient torque, and the previous system torque. The reward penalises deviation from the target angle, excessive velocity, high torque usage, strain, and abrupt changes in action.

This model captures the smooth and continuous behaviour expected from torque-driven actuation. Episodes terminate when the arm structure exits the safe range or when the maximum number of steps is reached.

\subsection{Simulation Behaviour}

Here we present the state transitions and the dynamics of simulation variables at each step. While the internal mechanisms differ, both environments compute the strain experienced by the patient. This strain serves as the feedback to the RL agent for minimisation.

\subsubsection{Kinematic Environment}

The kinematic environment represents the arm structure as a stepper-motor system, in which movement is controlled by discrete delays between motor steps. At each time step, the control input is a delay value $d_t$ (in milliseconds) chosen within the range of $[d_{\min}, d_{\max}]$. Smaller delays correspond to faster stepper rotation and, therefore, faster angular motion, while larger delays slow down the movement.

The angular progress of the arm structure is modelled as a fraction of the total number of motor steps:
\[
p_t = \frac{\text{current step}}{\text{max steps}}, \quad
\theta_t = p_t \cdot \theta_{\max}, \quad
\dot{\theta}_t = \frac{\Delta \theta}{d_t},
\]
where $\theta_{\max} = \pi/2$ is the maximum angle.

Forces acting on the joint include gravitational torque and inertial effects:
\[
\tau_g = m g l \sin(\theta_t), \quad
\tau_i = \tfrac{1}{3} m l^2 \cdot \frac{\Delta \dot{\theta}}{\Delta t}.
\]
The patient may apply a torque $\tau_p$, which can be modelled as either constant or sinusoidal to mimic different exercise profiles. The motor provides torque only when required, following an assist-as-needed rule:
\[
\tau_m = \max\Big(0, \, (\tau_g + \tau_i) - \tau_p\Big).
\]

To encourage safe and cooperative movement, a reward (or cost) signal is computed from five components:
\[
R_t = - w_{sp} E_{sp}
- w_{pb} E_{pb}
- w_{sf} E_{sf}
- w_{acc} E_{acc}
+ w_{pi} E_{pi}.
\]
Here, $E_{sp}$ measures deviation from a target speed profile, $E_{pb}$ penalizes situations where the patient blocks movement, $E_{sf}$ penalizes over-forcing by the system, $E_{acc}$ penalizes abrupt delay changes (encouraging smoothness), and $E_{pi}$ rewards active patient effort.

The episode ends when the maximum number of motor steps is executed. A rendering function visualises the arm's trajectory, illustrating how delay-based commands translate into movement.

\subsubsection{Dynamic Environment}

The dynamic environment provides a more detailed physics-based simulation. Here, the arm structure is modelled as a pendulum-like system subject to gravity, inertia, damping, and patient interaction. At each time step, the control input is the torque $\tau_s$ applied at the joint, constrained within $[0, \tau_{\max}]$.

The torques acting on the system include:
\[
\tau_g = - \tfrac{1}{2} m g l \sin(\theta_t), \quad
\tau_p = \text{patient torque}, \quad
\tau_f = - b \dot{\theta}_t,
\]
in addition to the applied torque $\tau_s$ and a patient-side gravitational torque $\tau_{pg}$. The total torque is therefore
\[
\tau_{\text{net}} = \tau_s + \tau_g + \tau_p + \tau_{pg} + \tau_f.
\]

This net torque determines the angular acceleration, which is integrated over the simulation time step $\Delta t$ to update velocity and position:
\[
\ddot{\theta}_t = \frac{3 \, \tau_{\text{net}}}{2 m l^2}, \quad
\dot{\theta}_{t+1} = \dot{\theta}_t + \Delta t \, \ddot{\theta}_t, \quad
\theta_{t+1} = \theta_t + \Delta t \, \dot{\theta}_{t+1}.
\]

In parallel, a hand-only trajectory $\theta_h$ is computed based on patient and gravitational torques. The difference between the system’s angle and the hand’s angle defines the strain:
\[
\Delta \theta = \theta_{t+1} - \theta_h,
\]
which reflects potential discomfort or misalignment between the patient and the device.

The reward function penalises physiologically undesirable behaviour as shown below:
\[
R_t = - \Big(
w_{\theta} |\theta_t - \theta^*|
+ w_v \dot{\theta}_t^2
+ w_{\tau} \tau_s^2
+ w_h \Delta \theta
+ w_{\Delta \tau} |\tau_s - \tau_{s,\text{prev}}|
\Big).
\]
The terms correspond, respectively, to deviation from the target angle, excessive velocity, high motor effort, strain between the patient and the system, and abrupt changes in torque.

Episodes terminate when the angle exceeds the safe range $[0, \theta^*]$ or when the system reaches the maximum number of steps. The environment also includes a termination cost equal to the system's velocity, suggesting that the policy should slow down gradually as it approaches completion.


\subsection{Training with Reinforcement Learning}

Of all the deep RL algorithms, Soft Actor-Critic (SAC) \cite{haarnoja2018softactorcriticoffpolicymaximum} was chosen
for several reasons.
The environments introduced here are new, which makes effective exploration essential. SAC includes an entropy term that allows the policy to explore broadly at the start of training and to reduce randomness gradually as the policy stabilises. This improves convergence and prevents the agent from collapsing into unsafe behaviours early on.

SAC is naturally suited for continuous control, aligning with the torque outputs in the dynamic environment and the delay values in the kinematic environment. Although SAC is more complex internally, inference is simple and efficient. SAC is also more sample-efficient due to its replay buffer. This is important for future real-world deployment, where other algorithms, such as Proximal Policy Optimisation (PPO)~\cite{schulman2017proximalpolicyoptimizationalgorithms}, require fresh on-policy data, making training impractically slow and data-hungry. The replay buffer enables SAC to repeatedly learn from past errors, an advantageous property for rehabilitation devices that must minimise mistakes. The separation between the critic operating in real-value space and the actor operating in a lower-precision, sparser control environment further improves stability.

Both simulators incorporate constraints that reflect real hardware behaviour. The kinematic environment offloads smoothness to the learning algorithm, whereas the dynamic model provides smoothness directly through physics. Together, they enable SAC to learn policies that are both effective and potentially transferable to real devices.

\subsection{Mapping the Simulators to the Hardware}

The real system accepts a desired next angle and the time required to reach that position. From these values, it returns the forces acting at the joint as feedback.

\paragraph{Kinematic model}
The kinematic actor receives the top force, bottom force, and current angle as inputs. It predicts the delay $d_t$, while the angular increment remains fixed:
\[
\theta_{t+1} = \theta_t + \Delta\theta.
\]
This matches the stepper-motor behaviour of the real hardware.

\paragraph{Dynamic model}
The dynamic actor receives the inputs: 
$
\left[\sin\theta_t,\ \cos\theta_t,\ \theta_t,\ \tau_{p,t},\ \tau_{s,t-1}\right],
$
and outputs a new system torque $\tau_{s,t}$. The hardware-side motion update is
listed as:
$
\ddot{\theta} = \frac{3\tau_{\text{net}}}{2ml^2}, \quad
\dot{\theta} = \dot{\theta} + \ddot{\theta}\Delta t, \quad
\theta_{t+1} = \theta_t + \dot{\theta}\Delta t.
$

\begin{figure}[ht!]
\centering
\includegraphics[width=1\linewidth]{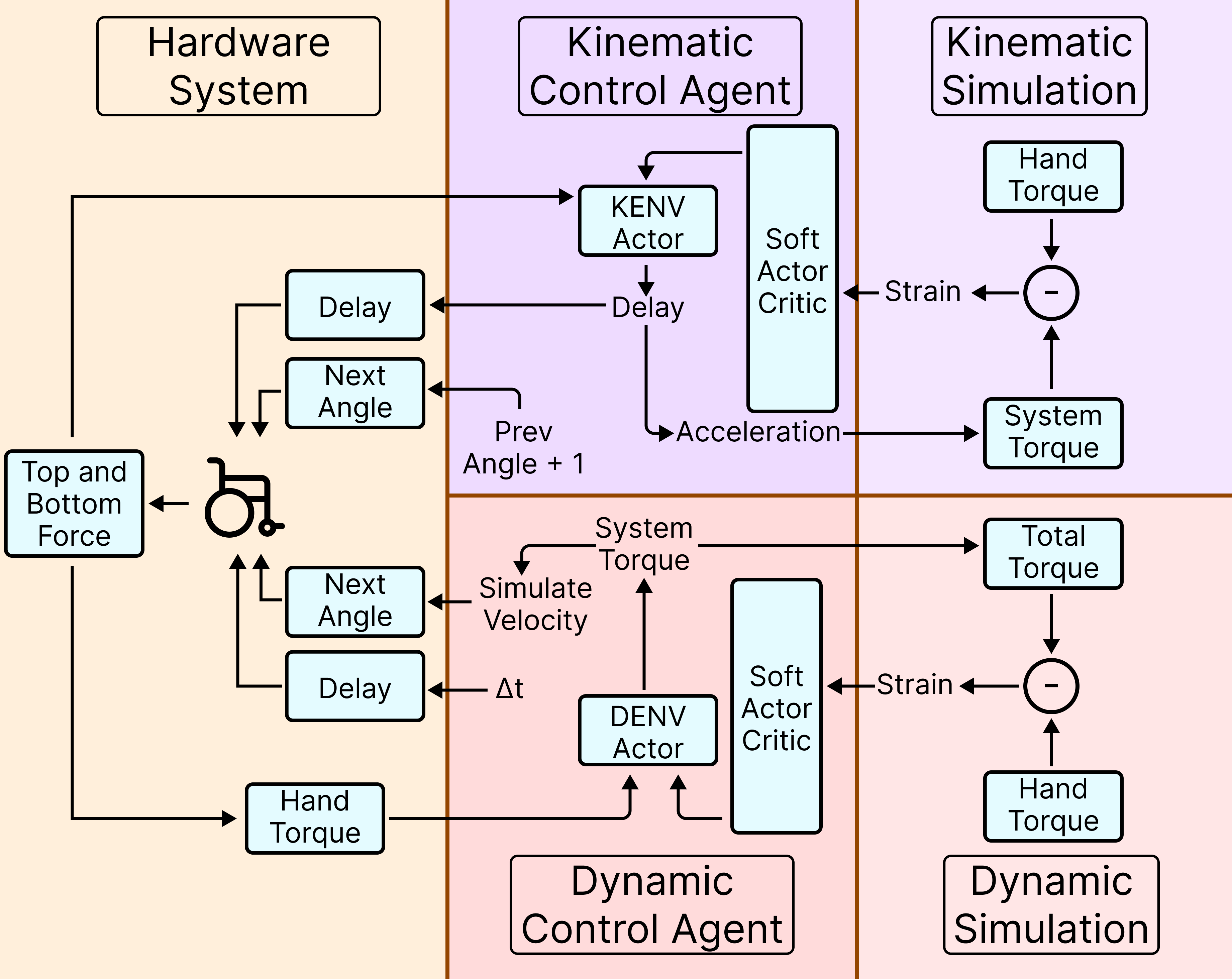}
\caption{Interaction between actors, simulations, and the actual hardware.}
\label{fig:act2hw}
\end{figure}

Patient torque is reconstructed from the following measured forces:
\[
\tau_{p} + \tau_{pg} = r F_{\text{net}}, \qquad
\tau_{pg} = -\frac{1}{2}mgl\sin\theta, \qquad
\tau_p = rF_{\text{net}} - \tau_{pg}.
\]
Thus, both simulation models map cleanly to the measurable and controllable quantities available on the physical device.
Figure~\ref{fig:act2hw} outlines the interactions between the simulation environments, the actor networks, and the actual hardware.

\subsection{Modifications to Reinforcement Learning Training}

Several adjustments were made to improve learning stability. First, the reward scale is intentionally not normalised to encourage the system to prioritise clinically desirable behaviours such as strain reduction. Second, the reward combines shaped terms with final progress terms, balancing local guidance with a clear terminal objective. These modifications helped the agent learn stable and clinically aligned behaviour in both simulation environments.


\section{Spiking Networks}

This section provides a preface for Spiking Neural Networks (SNNs) and the proposed optimisations. It consists of a background study, followed by an explanation of the hybrid training strategy and the inference tweaks tailored to building a regressor for continuous-valued control systems.

\subsection{Background}

\label{sec:snn-bg}

Spiking neural networks (SNNs) are a special form of artificial neural networks (ANNs) designed to mimic the functioning of neurons in the human brain. Research on SNNs provides increasing evidence of their greater latency and power efficiency compared to other architectures~\cite{snnReview-snnReview-yamazaki}, which aligns well with the constraints on the wheelchair. Spiking networks usually take the form of a typical ANN
with the only change being the addition of biological behaviour to the activation functions/neurons.
All neurons are modelled with a weighted connection between each pair of neurons. This connection weight is tuned using Spike-Time Dependent Plasticity (STDP): a process to alter the weight based on the temporal relationship between an input spike and an output spike.
Although the simplified mathematical SNNs drop time-based learning, they still retain the concept of weighted-spike transfer between neurons. From this, SNNs are formed by decomposing the weighted spike as the product of a weight scalar and a binary spike.
This section describes the components of an SNN used in our tests, including the neuron model, training and gradient-descent algorithms, encoders and decoders, the loss function, and any alterations.

\subsubsection{Neuron Models}

Various researchers replicated the dynamics of a biological neuron at varying levels of mathematical complexity and biological resemblance~\cite{neuronReview-typesOfNeurons-abusnaina}. These include Hodgkin-Huxley's and Morris-Lecar's, with the closest biological replication and high computational cost,
FitzHugh-Nagumo's model provides a balance, and the simplest adaptations, such as Leaky-Integrate-and-Fire (LIF) and Current-Based LIF(CUBA-LIF). Multiple attempts are also made to design neurons with custom dynamics tailored to a subset of use cases, like ALIF~\cite{learningToLearn-alif-bellec} and DALIF~\cite{dalif-dalif-phani}. This work uses the LIF neuron for its simplicity and flexibility to build SNNs.
At a high level, an LIF (Leaky) neuron
contains a hidden membrane potential that acts as its memory. The membrane voltage fluctuates with input activity and intrinsic time-based decay, and emits a spike when the membrane potential exceeds a particular threshold.

\begin{equation}
	S[t] = \begin{cases} 1, & \text{if}~H[t] \geq V_{\rm th} \\
              0, & \text{otherwise}
	\end{cases}
	\label{eq:lif-spiking}
\end{equation}

\begin{equation}
	H[t+1] = \beta H[t] + WX[t+1] - S[t]V_{\rm th}
	\label{eq:lif-updateHidden}
\end{equation}

\begin{equation}
    H[t+1] = (\beta H[t] + WX[t+1])(1 - S[t])
    \label{eq:lif-zeroReset}
\end{equation}

Equations \ref{eq:lif-spiking} and \ref{eq:lif-updateHidden} represent the mathematical transition dynamics of the LIF neuron. Drawing similarities with the matrix representation from an ANN,

\begin{itemize}
	\item $H[t]$ and $S[t]$ represent a neuron's membrane potential and spike output at time step $t$
	\item $\beta$ represents the membrane voltage decay multiplier
	\item $W$ denotes the STDP weight of a neuron layer
    \item $V_{th}$ denotes the spiking threshold voltage.
\end{itemize}
The weight matrix $W$ and decay $\beta$ are set to be trainable during supervised learning. Equation~\ref{eq:lif-spiking} is a shifted Heaviside function and represents the spiking activity as a function of membrane voltage. Based on the spike activity, the internal neuron state updates according to Equation~\ref{eq:lif-updateHidden}. This uses the subtract reset mechanism, which leaks $V_{th}$ volts to generate a spike. Equation~\ref{eq:lif-zeroReset} shows the zero reset method, which resets the membrane when a spike occurs.

\subsubsection{Data Encoding and Decoding}

An SNN takes data in multiple ways, typically after an encoding process.
Examples of these include rate coding
and temporal coding
~\cite{neuralCodingReview-encoding-guo}. Despite their popularity, these methods require knowledge of the range of values and would also remove the significance of the relative scaling of input values. Since our environmental input is relatively dynamic, we use direct encoding, which immediately feeds the input value to the network for a fixed number of time steps, $T$. Nitin et al.~\cite{dietSNN-directEncoding-nitin} investigated direct encoding in the context of image classification and achieved respectable levels of accuracy with major gains in latency and power.

Similar to the input, SNN output spikes must be decoded into a real value to work 
as a regressor. Rate and temporal decoding methods exist to complement their encoding counterparts, but they primarily target probability predictions.
For direct encoding, a straightforward way is to take a linear combination of output spikes. For this work, we use the population-based Output Weighted Spike decoding scheme as proposed in 
\cite{spikeDecoders-ows-phani}. Following equation~\ref{eq:ows}, this method takes a weighted average of spikes from a population of output neurons at the end of the forward pass.
This is denoted by lines 7 and 10 in pseudocode~\ref{alg:decoders}. Rate decoding, on lines 6 and 9, stores all the output spikes, which requires more memory to perform the relevant decoding operation on the array.

\begin{equation}
    \hat{Y} = \sum_{i=0}^{\kappa}W_{i}*X_{i}[T] + b
    \label{eq:ows}
\end{equation}

\subsubsection{Loss Function}

Spiking networks are often trained with customised loss functions to accommodate the time dimension. These loss functions also usually complement the encoder-decoder combination of the network, like mean squared rate error (MSE Rate Loss), cross entropy count losses, and other statistical measures~\cite{snntorch-lossfunctions-jason, spytorch-cemembraneloss-friedemann, slayer-mseCountLoss-sumit}. Converting the spike train to the real-value domain enables compatibility with distance and value-based loss functions. And in the case of the SAC algorithm, the OWS decoder adapts the SNN-based actor network to the original MSE loss function.

\begin{algorithm}
\caption{Forward Pass for Each Input $x$ with Rate and OWS Decoders.}
\label{alg:decoders}
\setlength{\commentlen}{22ex}
\begin{algorithmic}[1]
\State \textbf{Input:} x$\in \mathbb{R}^{[n]}$, $T$, $\phi$
\State \textbf{Init:} $S$ = [] \Comment{\makebox[\commentlen][l]{Array to record Spikes}}
\State \textbf{$X$} $ \in \mathbb{R}^{[T,n]} \gets \mathbb{E}_{rate}(x)$ \Comment{\makebox[\commentlen][l]{Rate Encoding}}
\For{$t$=0 to $T$ time steps}
    \State $x^{rate}_t \gets \phi(X[t])$
    \State $S \gets S \cup x^{rate}_t$ \Comment{\makebox[\commentlen][l]{Store Output Spikes}}
    \State $x^{ows}_t \gets \phi(x)$ \Comment{\makebox[\commentlen][l]{Direct Encoding}}
\EndFor
\State $\hat{y}_1 \gets \mathbb{D}_{rate}(S)$ \Comment{\makebox[\commentlen][l]{Rate Decoder}}
\State $\hat{y}_2 \gets \mathbb{D}_{ows}(x^{ows}_T)$ \Comment{\makebox[\commentlen][l]{OWS Decoder}}
\State $l_{rate} \gets CCE_{rate}(\hat{y}_{1}, y)$ \Comment{\makebox[\commentlen][l]{Special loss fn}}
\State $l_{ows} \gets MSE(\hat{y}_{2}, y)$ \Comment{\makebox[\commentlen][l]{Standard MSE function}}
\end{algorithmic}
\end{algorithm}


\subsubsection{Gradient Descent}

SNNs use a surrogate gradient system to avoid the Vanishing Gradient problem~\cite{surrogateGD-surrogate-emre}. This occurs when gradients are computed from binary values during backpropagation.
In our paper, we use the fast sigmoid function as a close, continuous mathematical approximation to the spiking Heaviside function in just the backward pass.
The forward equation approximating a spike pulse and its first derivative are presented with the equation set~\ref{eq:lif-backward}, where $k$ represents the slope/curve tightness.

\begin{equation}
	\begin{split}
		S[t] & \approx \frac{H[t]}{1 + k|H[t]|} \\
		\frac{\partial S[t]}{\partial H[t]} & = \frac{1}{(1 + k|H[t])^2}
	\end{split}
	\label{eq:lif-backward}
\end{equation}

Using the computed gradients, the network parameters are trained using the Back Propagation Through Time (BPTT) algorithm~\cite{superspike-bptt-friedemann}, which performs gradient descent across the network structure and through time.
The membrane potential $H$, membrane decay $\beta$, and membrane threshold $V_{th}$ of the LIF neurons are involved in the forward computation. This enables the gradient-descent-based BPTT algorithm to optimise $\beta$ and $V_{th}$, thereby allowing the network to learn the task at the neuron level. There have been successful attempts to train an SNN using evolutionary methods~\cite{evotraining-evomethods-daniel, evotrainStrategy-evomethods-shen, autoracing-evomethods-patton}. We used BPTT for a mathematically predictable and replicable training curve on the untested environments.

This paper proposes two optimisations for SNNs over the State-of-the-Art methods for SNNs: 1. Heterogeneous SNN Architecture targeting the training phase, 2. Temporal quantisation with continuous thought for the inference stage.

\subsection{Heterogeneous SNN Architecture}

To optimise power usage on the edge machine, we designed the hardware system to include a neuromorphic chip for inference. Since only the actor network of the SAC algorithm is active during inference, we propose training the actor network directly in the neuromorphic domain, while retaining the critic in the artificial domain. This way, we theorise that retaining a part of the algorithm in higher precision leads to better learning.

Several authors have proposed a neuromorphic RL actor network for edge computing. Patel et al.~\cite{convertedSNNrl-conversion-patel} proposed converting an ANN-based Deep Q-learning algorithm to an equivalent SNN with weight-proportional neuron characteristics. When tested on Atari Breakout, the authors found that converted SNNs outperformed standard ANNs. While they suggest that SNNs built this way can adapt better to unseen states, the main drawback remains the need to run the SNN for 500 time steps, which induces higher latency and power consumption.

Another way is to train the network in a fully spiking domain
as suggested by Junqi et al.~\cite{actorCriticSNN-sacSS-junqi}. The authors tested the actor-critic algorithm on various control tasks using a complete SNN setup with temporal (time-to-first-spike, TTFS) encoding. The authors claimed a 96\% success rate in obstacle avoidance when controlling an Unmanned Aerial Vehicle (UAV) and insist that a temporally coded actor-critic SNN is suitable for control applications. Although other results from their paper support that argument, implementing this on high-precision systems may lead to subpar performance. Our work considers the Soft-Actor-Critic (SAC)- adapted version of the above (Spiking-SAC, SSAC) as the State-of-the-art (SOTA) baseline for the spiking domain.


Authors in \cite{popsan-3network-tang} proposed a more direct architecture for actor-critic networks. They proposed using one SNN to predict the mean $\mu$, and an ANN to predict the variance $\sigma$ forming the actor component, and one ANN to predict the state-action value $Q_{(s,a)}$ as the critic.
The authors also employed population decoding with rat-encoded input. While performing a weighted sum on spike rate provides high precision with promising results, as discussed in 
~\cite {spikeDecoders-ows-phani}, rate decoding adds latency at the input and output of the network during the spike accumulation stage.

Hence, we propose a novel Hybrid-SAC (HSAC): a new variant with algorithmic tweaks and incorporating the highlights of SOTA. HSAC uses OWS decoding and direct encoding for speed and a single SNN to predict both the mean and variance. Since we use OWS, SNN outputs standard floating values, enabling us to use Mean Squared Error (MSE) as the loss function.
Algorithm~\ref{alg:HSACGrad} provides the steps involved in gradient-based learning with two different network architectures. Compared to standard SAC, no changes are necessary, since the loss is applied to a decoded real value rather than a spike train.

\begin{algorithm}
\caption{HSAC Gradient Learning}
\label{alg:HSACGrad}
\begin{algorithmic}[1]
\State ...
\For{each environment step}
    \State $a_t \gets \mathbb{D}_{ows}(\pi_\phi(a|s_t))$
    \State $s_{t+1}, r_{t} \gets env.act(a_t)$
\EndFor
\State Explore and Store Interactions
\For{each gradient step}
    \State $\theta \gets \theta - \lambda_Q \nabla_{\theta} J_Q(\theta)$ \Comment{Update ANN Critic Net}
    \State $\phi \gets \phi - \lambda_\pi \nabla_\phi J_\pi(\phi)$ \Comment{Update SNN Actor Net}
    \State $\alpha \gets \alpha - \lambda_\alpha \nabla_\alpha J(\alpha)$ \Comment{Update Entropy Factor}
\EndFor
\State ...
\end{algorithmic}
\end{algorithm}

Although this reduces precision compared to the rate-population decoder, OWS compensates for the loss by providing a trainable set of floating-value weights for each output. Overall, our proposed network architecture achieves the following:

\begin{itemize}
    \item increases training efficiency by decreasing the total number of neural networks to two compared to POPSan~\cite{popsan-3network-tang},
    \item reduces the time-steps drastically compared to Patel et al.~\cite{convertedSNNrl-conversion-patel},
    \item employs appropriate architecture for better use of available hardware compared to Spiking only SAC~\cite{actorCriticSNN-sacSS-junqi}
\end{itemize}

\subsection{Quantized Inference}

This subsection proposes the Spiking Post-Training Temporal Quantisation (SPTTQ) method and the accompanying Sequent Leaky ({\it SLeaky}) neuron model for latency- and spike-efficient prediction in continuous RL.

\subsubsection{SPTTQ}
Spiking networks require performing the forward pass over multiple time steps using the same input value, effectively encoding the input into a data type compatible with SNNs.
Usually, the number of time steps, as a hyperparameter, remains constant during training and inference. The network then learns to emit the correct number of spikes only by the end of a forward pass, as determined by rate decoders.
Since each spike requires energy, power optimisation techniques to reduce spike activity are explored.


One such method proposes adding spike activity as a learnable target to supervised learning~\cite{sinabs-spikeactivityloss-sheik, nrgOptSNN-spikeactivityloss-sorbaro}. Spike activity loss is defined as the difference between the total spikes emitted
and the desired spike count/rate. The higher the loss, the more the network trains to reduce per-neuron spike activity and, by extension, makes the network sparsely spiking. Although this helps reduce power consumption, it does not completely address the latency problem, since the network still has to process all time steps.

Sayeed et al.~\cite{tempPruning-pruning-sayeed} presented another methodology called temporal pruning, where time steps are removed during training
until the network fails to learn. They investigated this technique on ImageNet classification and 
achieved a competitive performance compared to their baselines.
Although this reduces latency, it forces the network to learn with higher spike activity to compensate for the reduced computation. This method also requires training the SNNs several times, which increases the computational requirement. Yuhang et al.~\cite{seenn-earlyexit-yuhang} also investigated a similar approach by the name Spiking Early Exit Neural Nets (SEENN). Specifically, SEENN-I, as described in their work, is a post-training technique that uses a confidence score, proportional to the prediction class probability, to exit early. While they exhibit higher accuracy even at lower time steps, the concept of a confidence score relates to the logic of classification probabilities and cannot be directly applied to regression. An additional branch to the SNN's output is necessary to adapt this calculation to a regression network.

This is where our proposed methodology fills the gap, by defining an early exit strategy for regressive and control tasks, called Spiking Post-Training Temporal Quantisation (SPTTQ), designed for control applications. The gist of SPTTQ is to evaluate a trained SNN in the environment over multiple episodes with various cut-off steps, ranging from the maximum time steps down to 1. Once benchmarked, we statistically pick the time step that offers a balance between spike activity and performance to be the designated short time step $\tau$. The network then runs for these many steps in inference rather than the complete $T$ time steps. The mechanism used for picking $\tau$ can be one of: the plateau point when the performance starts to drop, or when the variance of output values across time steps exceeds a designated threshold. By doing so, we are banking on the SNN to reach a stable output value at a time step earlier than 
expected.

This early on-par performance is achieved by the OWS decoder. The decoder trains an SNN to predict the correct number of spikes only on the last time step. Yet, as observed in the behavioural results in Section~\ref{sec:res-main}, learning happens linearly due to BPTT and the involvement of neuron variables. Thus, the OWS decoder ensures that neurons reach the required spike activity at the earliest time step. SPTTQ exploits this quirk to determine an optimal time-step cut-off for a given trained SNN+OWS combo. As with standard post-training quantisation techniques for ANNs, SPTTQ operates only on trained SNNs. Hence, i) a time step reduction must not be expected all the time due to the varying complexity of the environment,  
ii) The SNN training process 
need not be altered and requires the network to train for all the designated time steps $T$, and 
iii) the network need not be retrained, since this removes computationally redundant operations while retaining most of the performance.

Hence, to summarise, SPTTQ saves power and latency in a trained SNN by:
\begin{itemize}
	\item reducing the time steps instead of just focusing on spike activity like in the SinABS Example~\cite{sinabs-spikeactivityloss-sheik} and Sorbaro et al.~\cite{nrgOptSNN-spikeactivityloss-sorbaro}
	\item not retraining the network repetitively to increase spike activity and gain accuracy as in Temporal Pruning~\cite{tempPruning-pruning-sayeed}
	\item Compute the exit iteration in compile time to save the additional floating point operation latency in runtime induced by SEENN-I~\cite{seenn-earlyexit-yuhang} while adopting a similar structure for the regression problem.
\end{itemize}

\subsubsection{Sequent Leaky Neuron}

\begin{figure}[ht!]
    \centering
    \includegraphics[width=1\linewidth]{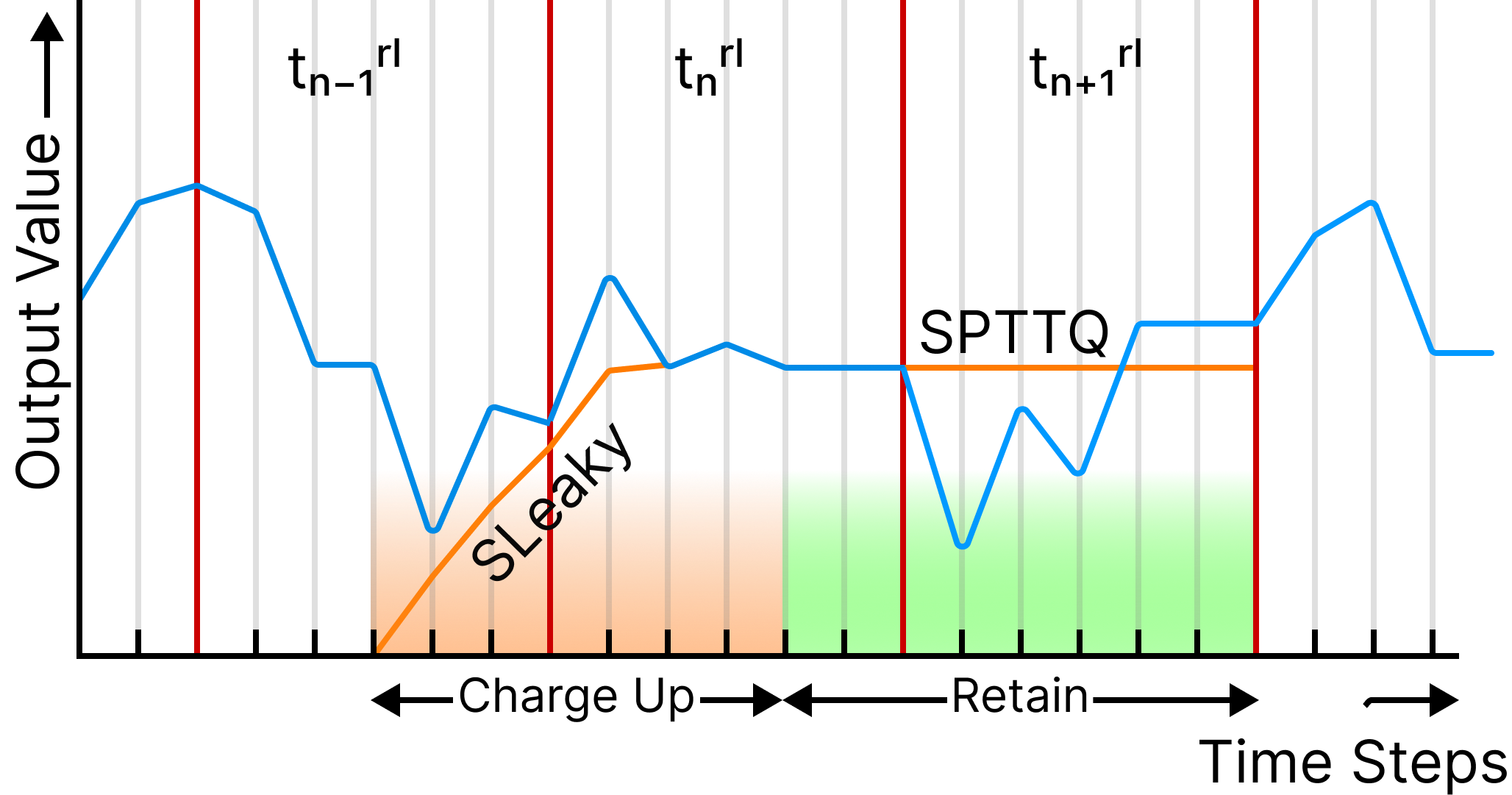}
    \caption{Cut-off regions introduced by SPTTQ and {\it SLeaky}'s no-reset functionality. Without them, the original orange path requires 16 steps to complete one RL step. The optimised path, represented by blue lines, shows faster transitions between RL steps. Also shows the charge-up phase in orange and the retain phase in green within one RL step.}
    \label{fig:snnEffects}
\end{figure}

While SPTTQ trims the end of the time-steps, 
further latency savings can be achieved at the start of the forward pass.
Neurons in an SNN reset to zero membrane potential after one forward pass, equivalent to flushing all the charge within. Logically, this will disconnect the state information of one forward pass from the next pass, which is typically necessary for classification problems. The neurons should then recharge to emit the required number of spikes for the next input. This process from zero voltage to the desired level
is further divided into two phases: i) Charge-up phase, and ii) Retain phase. These phases are highlighted in orange and green shades in Figure~\ref{fig:snnEffects}, respectively. As the name suggests, the charge-up phase builds up the membrane potential of all the neurons from zero, while the retain phase maintains the membrane potential. Resetting the neurons to zero will necessitate a charge-up phase for the next input. As an enhanced approach, we propose retaining the membrane potential between forward passes until the end of an episode. Since the output is expected to change only moderately in some RL environments, retaining the membrane potential will require only a mathematically gradual change in neuron states, and, by extension, will require fewer time steps for the next charge-up phase.

Authors in Hwang et al.~\cite{preChargedMem-precharging-hwang} proposed a method called Pre-Charged Membrane Potential (PCMP). This technique charges the neurons by a fixed amount at the start of a forward pass.
The authors evaluated this using multiple networks with ANN-SNN conversion
and rate decoding to perform image classification.
With PCMP, the authors observed SNNs to converge
faster as the precharge voltage increased.

To avoid additional logical steps to reset and pre-charge, we never reset the neuron state to achieve a similar effect. Pre-charging effectively pushes the neurons ahead in time and skips the charge-up phase, thereby reducing latency. This differs from the cited method in using old cell-state copies and varying voltage values for each neuron, whereas authors in Hwang~\cite{preChargedMem-precharging-hwang} proposed only optimised constant precharge values at the network and layer levels. Hence, this supports our claim that retaining the charge between RL steps helps to reduce the network's forward-pass latency.

To accommodate the new behaviour, we propose the Sequent Leaky ({\it SLeaky}) neuron model. This adds a pseudo-detach reset method for inference only and negative-threshold checking. The new reset functionality includes detaching the hidden states of a neuron at the end of a forward pass and copying them to the initial states 
for the following input. Algorithm~\ref{alg:noResetInf} presents the flow of the no-reset/copy-forward method that is employed during the inference run. The network retains the membrane voltage after each forward pass by calling the $continue$ function on line 5 and resets only after an episode, as indicated on line 10. Figure~\ref{fig:snnEffects} presents an example scenario of time steps saved as represented by the orange line at the start for the RL step $t_{n}^{rl}$.

Since the SNN does not reset during an episode, we must solve any potential side effects.
Firstly, negative signals due to negative weights in the network usually slip past the standard LIF step, as modelled in Equation~\ref{eq:lif-updateHidden}, and cause the network to deteriorate into a non-spikable state. The basic LIF implementation doesn't constitute membrane voltage clamping, allowing SNNs to learn as ANNs do. Since a complete reset occurs only at the end of an episode, a highly negative membrane will take lots of time steps to recover from a non-spikable state. Hence, the new continue method also clears any negative voltages. This will not affect the neuron dynamics, since only positive signals are propagated past a neuron. Next, we use the zero-reset mechanism as the post-spike behaviour. The subtract-based spike reset mechanism in an LIF neuron will leak the voltage by the threshold amount $V_{th}$ if a spike is emitted. Since we train the SNN using deep learning methods and the voltage never resets the network till the episode ends, any occurrence of a high weight value will cause the membrane voltage to rise, leading to continuous spiking.
Zero reset, defined by Equation~\ref{eq:lif-zeroReset}, will completely drain the cell and keep the voltages within range.
Although the zero-reset method contradicts the notion of saving power by flushing all the charge, the latency and power reduction provided by the SPTTQ+{\it SLeaky} combination outweigh any potential power savings with the subtract-based spike reset mechanism. Plus, the zero-reset method avoids the risk of runaway membrane voltage.


\begin{algorithm}
\caption{Inference loop for each episode}
\label{alg:noResetInf}
\setlength{\commentlen}{15ex}
\begin{algorithmic}[1]
\State \textbf{Input: } $env, N, \phi'$
\State \textbf{Init: } $x \gets env.reset(), R \gets 0$
\For{$i$=0 to $N$ RL Steps}
    \State $a \gets \phi'(x)$  \Comment{\makebox[\commentlen][l]{Forward pass}}
    \State $\phi'.continue()$  \Comment{\makebox[\commentlen][l]{Carry forward call}}
    \State $(r, x') \gets env.perform(a)$   \Comment{\makebox[\commentlen][l]{Perform action}}
    \State $x \gets x'$
    \State $R \gets R + r$
\EndFor
\State $\phi'.reset\_mem()$  \Comment{\makebox[\commentlen][l]{Membrane reset}}
\end{algorithmic}
\end{algorithm}

\subsubsection{Combination}
In a nutshell, the proposed changes primarily aim to optimise performance, runtime, and power.
This is done by using suitable hardware architectures that meet the environment's constraints using HSAC, conserving power-state information over time using {\it SLeaky} neurons, and statistically quantising the runtime system in the time domain using SPTTQ.
As a result of eliminating redundant operations, the forward pass is computationally optimised.
To compensate for the missing charge at the start of an RL episode, the first RL step also runs for all $T$ time steps to allow the network to reach an optimal spiking state. This charge is then held for the entire RL episode.
These optimisations are designed to integrate harmoniously with the RL algorithm, so they self-correct shortcomings at each RL step, providing flexibility to scale the tweaks.

Pseudocode referred in Algorithm~\ref{alg:optSNN} lists the final procedure to implement the above optimisations. The network optimiser takes the trained SNN \& a minimum performance threshold as inputs, and returns a new SNN with {\it SLeaky} neurons, with a new short time step $\tau$.

\begin{algorithm}
\caption{Optimize SNN Actor}
\label{alg:optSNN}
\begin{algorithmic}[1]
\State \textbf{Input:} $\phi$, $\delta$, $T$
\State $r_b \gets eval(\phi, T)$
\State $\phi' \gets \phi[LIF \to SLIF]$
\For{$t$=$T$ to 1, step = -1}
    \State $r_b^t \gets eval(\phi, t)$
    \If{$r_b^t \le \delta*r_b$}
        \State $break$
    \EndIf
\EndFor
\State \textbf{Output: } $\tau, \phi'$
\end{algorithmic}
\end{algorithm}

\sectionEND{Spiking Networks}

\section{Experiments, Results and Discussion}
\label{sec:res-main}
\begin{figure*}[ht!]
    \centering
    \includegraphics[width=1\linewidth]{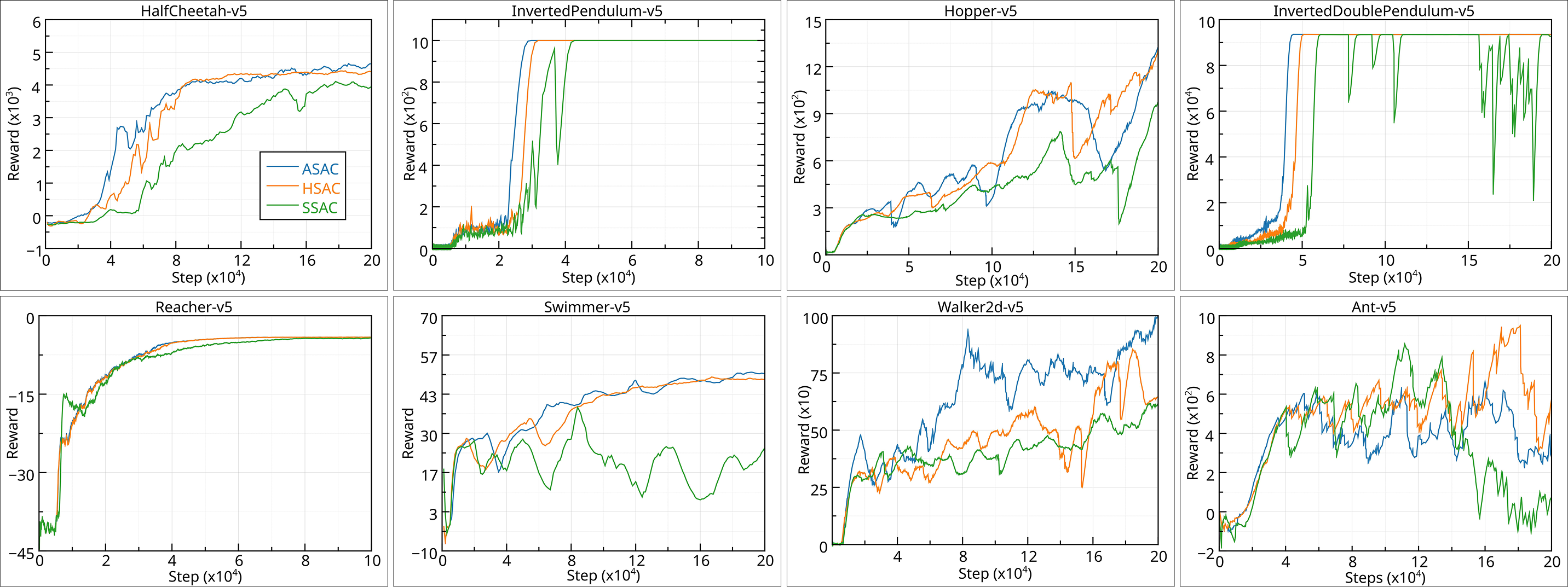}
    \caption{Training Reward versus Global Steps plot for 8 MuJoCo Environments.}
    \label{fig:mujocoHSAC}
\end{figure*}
This section discusses all the experiments performed and results obtained with the proposed designs.
We primarily evaluated the performance of the proposed tweaks and assessed their behaviour against SOTA methods. This involves testing the HSAC method on a subset of MuJoCo, KENV, and DENV environments. Later, we present the behavioural study of SPTTQ and the No-Reset-based Sequent Leaky neuron, followed by a comparison with prior-art methods. The SOTA methods are reproduced in our testing environment to ensure a fair comparison of the learning algorithms. Finally, we present the effects of the reward function on simulation behaviour and the actor policy to understand its components' weightage. Appendix section~\ref{app:hparams} includes all the hyperparameters used in these tests.


\subsection{Heterogeneous Trainings}



We compare the new HSAC algorithm with the pure spiking form, Spiking SAC (SSAC), and the artificial variant, Artificial-SAC (ASAC). We consider ASAC and SSAC as baselines in their respective domains for comparison with HSAC.

Initially, the hybrid training strategy is evaluated on the reference MuJoCo environments to validate the algorithm. A relatively deep, large network structure was chosen for all tests to ensure convergence even with complex environments. The training loop was limited to 5 million steps, with an actor learning rate (LR) of 1$e^{-4}$ and a critic LR of 3$e^{-4}$. Other hyperparameters are provided in Appendix~\ref{app:hparams}.

\begin{figure}[ht!]
    \centering
    \includegraphics[width=1\linewidth]{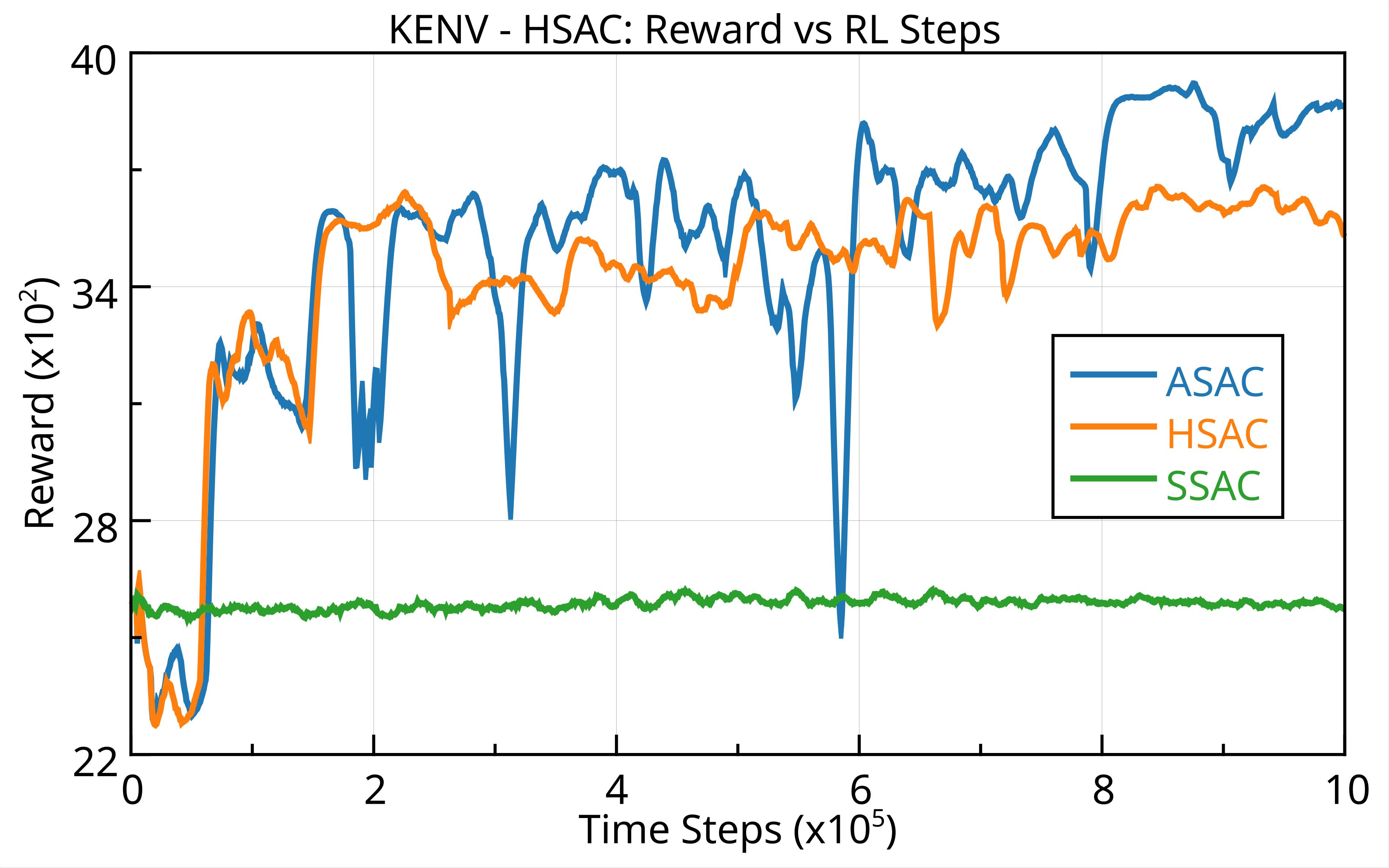}
    \caption{Training Reward versus Global Steps plot for Kinematic Environment.}
    \label{fig:kenvTrain}
\end{figure}

\begin{figure}[ht!]
    \centering
    \includegraphics[width=1\linewidth]{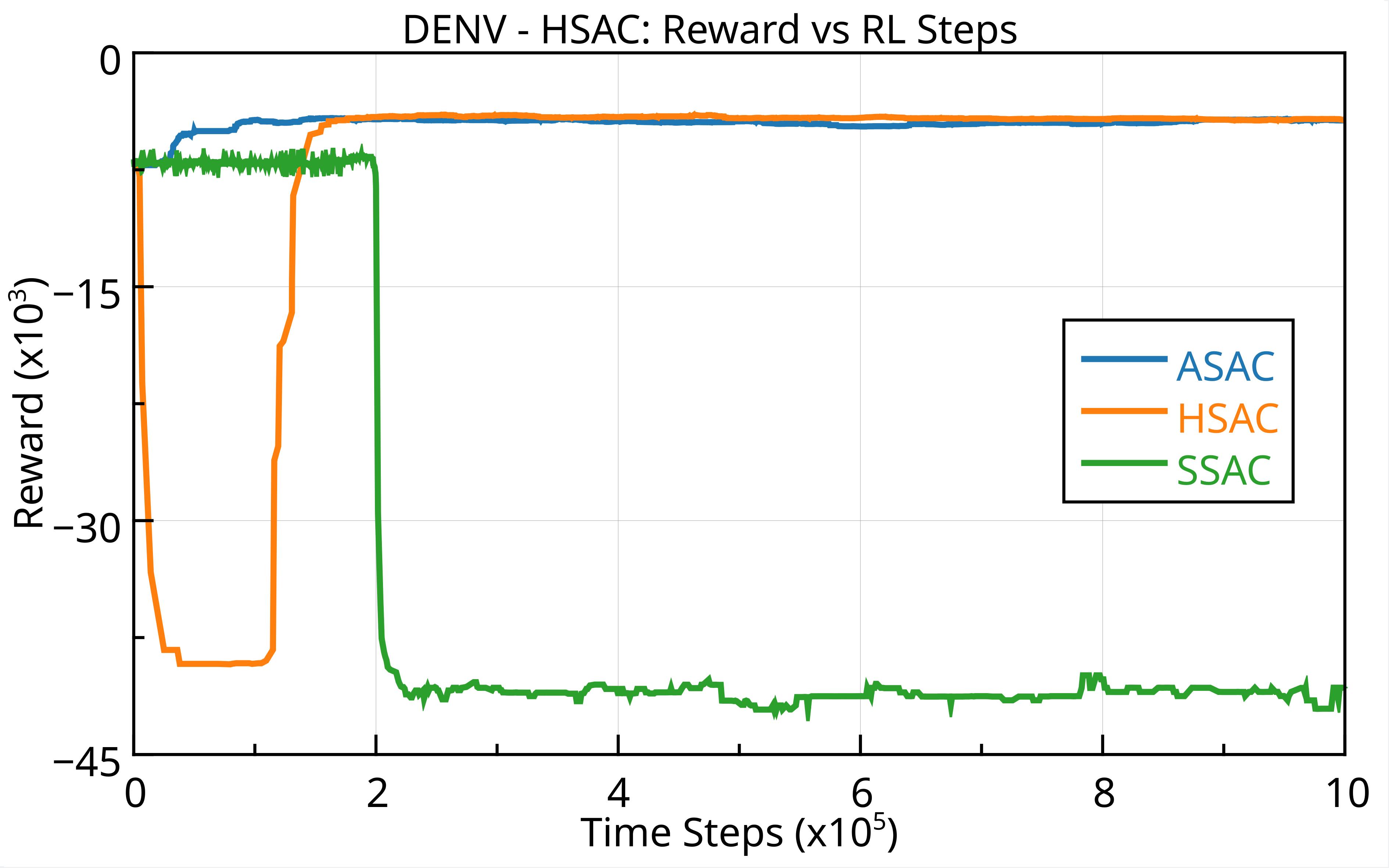}
    \caption{Training Reward versus Global Steps plot for Dynamic Environment.}
    \label{fig:denvTrain}
\end{figure}

Plots in Figure~\ref{fig:mujocoHSAC} show the training history leading to convergence across 8 MuJoCo environments. Each graph plots the episodic total reward $R$ against the global RL step. The ASAC algorithm learned the objective in 6 of the 8 environments, but failed to learn the objectives of Reacher and Swimmer. We also observed that HSAC outperformed SSAC across all metrics, while being slightly slower than ASAC. Environments such as HalfCheetah and InvertedDoublePendulum showcase the best flexibility of HSAC. HSAC provided the stability of full-precision ASAC training, yet yielded a hardware-specific neuromorphic network, similar to the less stable and slower SSAC. Hence, this validates the HSAC method.

A similar behaviour is observed when we trained HSAC on DENV and KENV environments. The reward versus RL steps data in DENV \& KENV environments are shown in Figure~\ref{fig:denvTrain} and Figure~\ref{fig:kenvTrain}, respectively. While the SSAC algorithm completely fails to learn the environments, HSAC trains successfully. Even though it learns more slowly than ASAC, it eventually catches up.

\subsection{Inference Behaviour}
This section provides a behavioural study of the novel methods for performing SNN inference. KENV is used as the testbed for all experiments in this study.

\begin{figure}[htb!]
    \centering
    \includegraphics[width=1\linewidth]{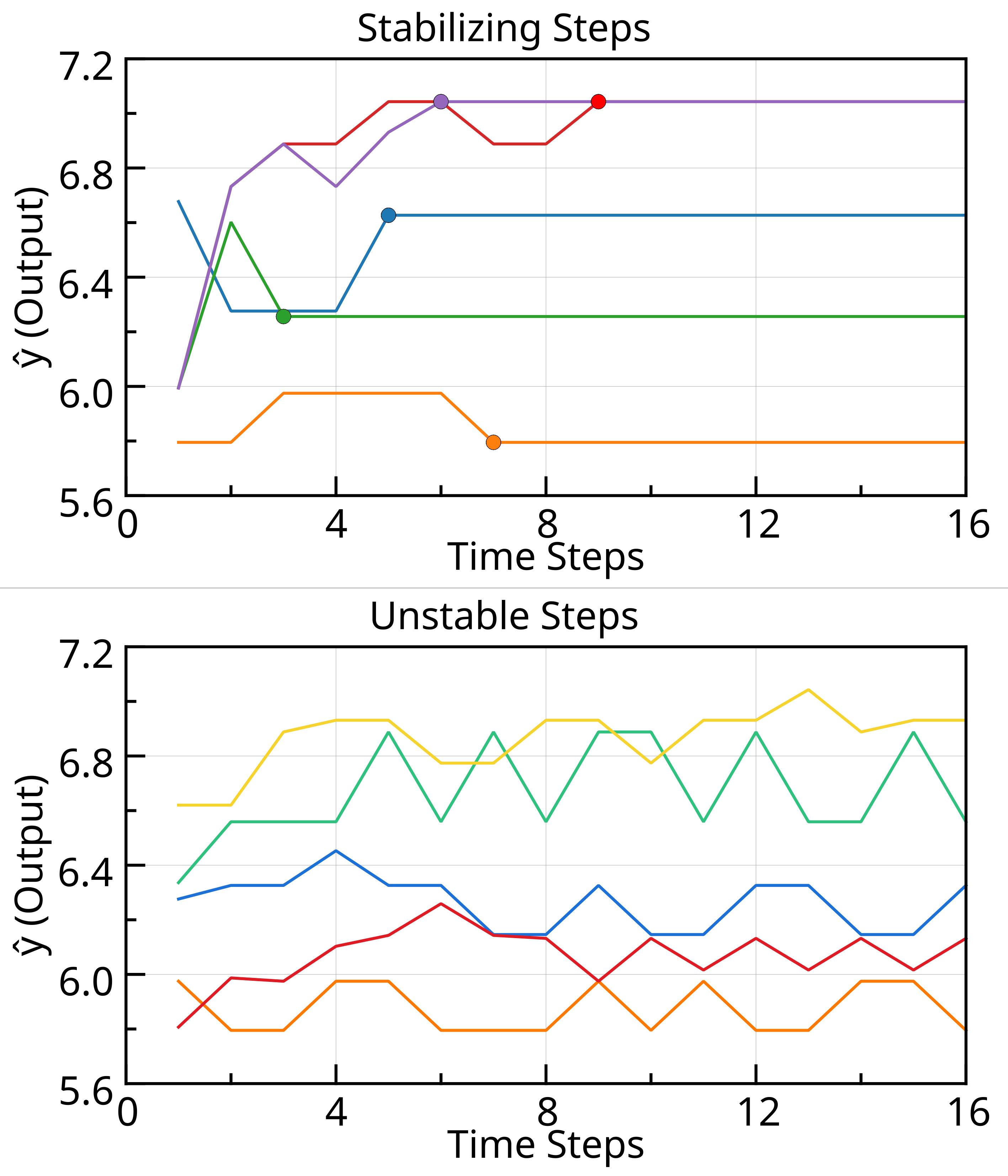}
    \caption{Examples of stable and unstable RL steps, with the onset of flattening marked for stable cases.}
    \label{fig:lines}
\end{figure}

First, we show the effect of the OWS decoder on output values across time steps. Based on the changes in the output,
an RL step in the spiking domain is divided into two phases: stable and unstable. An RL step is deemed stable if the decoded predicted value at each time step stays constant for some time steps till the end. 
These are the best examples to show 
the effect of OWS on SNN to learn quickly across the time domain.
However, some samples exhibited constant fluctuations in the output value, which are
referred to as unstable RL steps.
Figure~\ref{fig:lines} shows five stable and unstable RL steps. The time steps until the marked circle for stabilising lines are the charge-up phase, and beyond that is the retain phase. Unstable points continue to fluctuate until the end, and hence, don't have any marked circles.

\begin{figure}[htb!]
    \centering
    \includegraphics[width=1\linewidth]{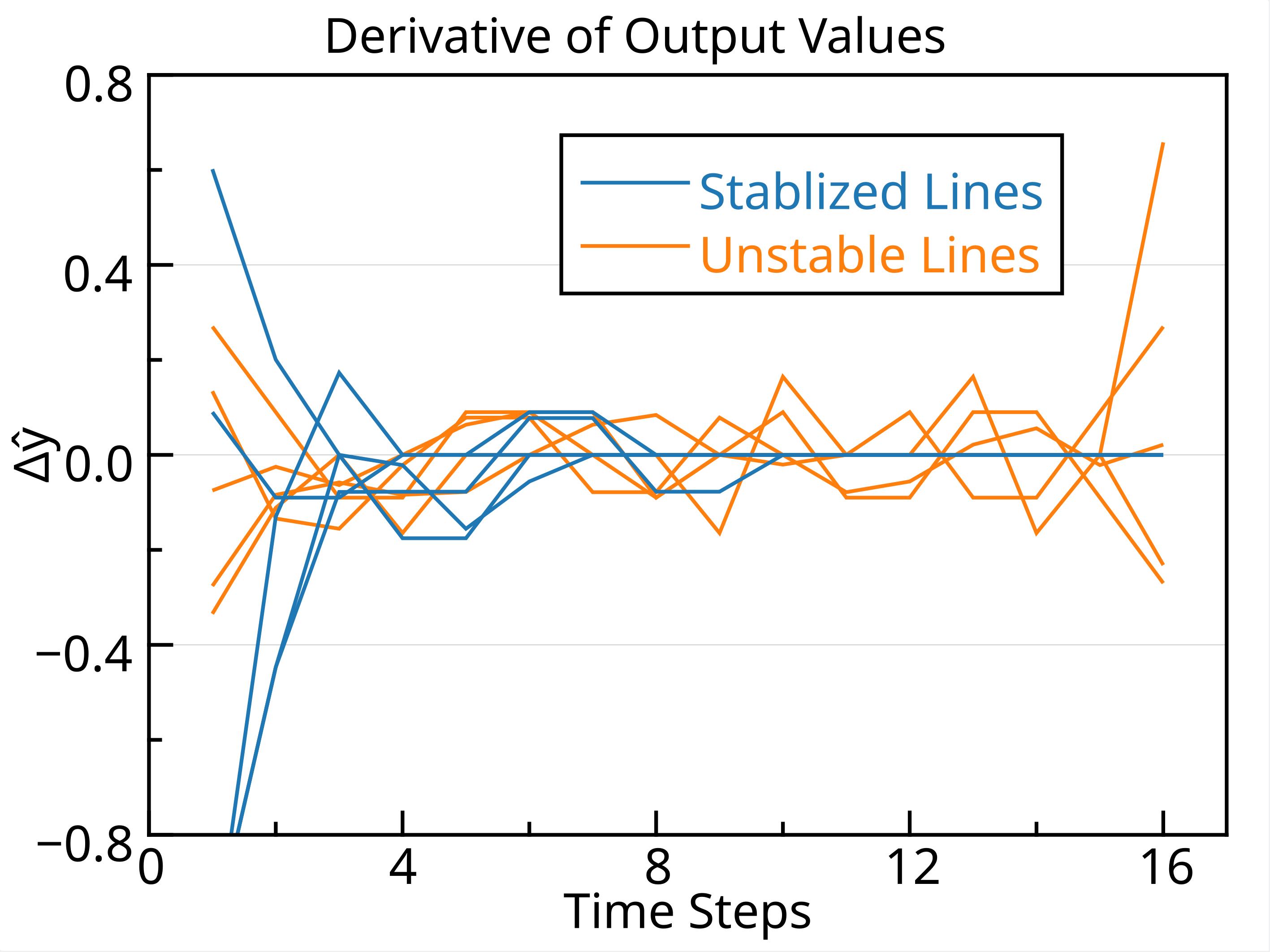}
    \caption{Derivatives of stable and unstable outputs in which stable trajectories converge to zero while unstable ones continue fluctuating.}
    \label{fig:derivative}
\end{figure}

Figure~\ref{fig:derivative} plots the derivative of the output value. We observe that the derivative of stable lines collapses to zero at some time, while the derivative of unstable points never stabilises. SPTTQ benefits from these stable points because the faster the network reaches them, the more steps it can skip.

\begin{figure}[htb!]
    \centering
    \includegraphics[width=1\linewidth]{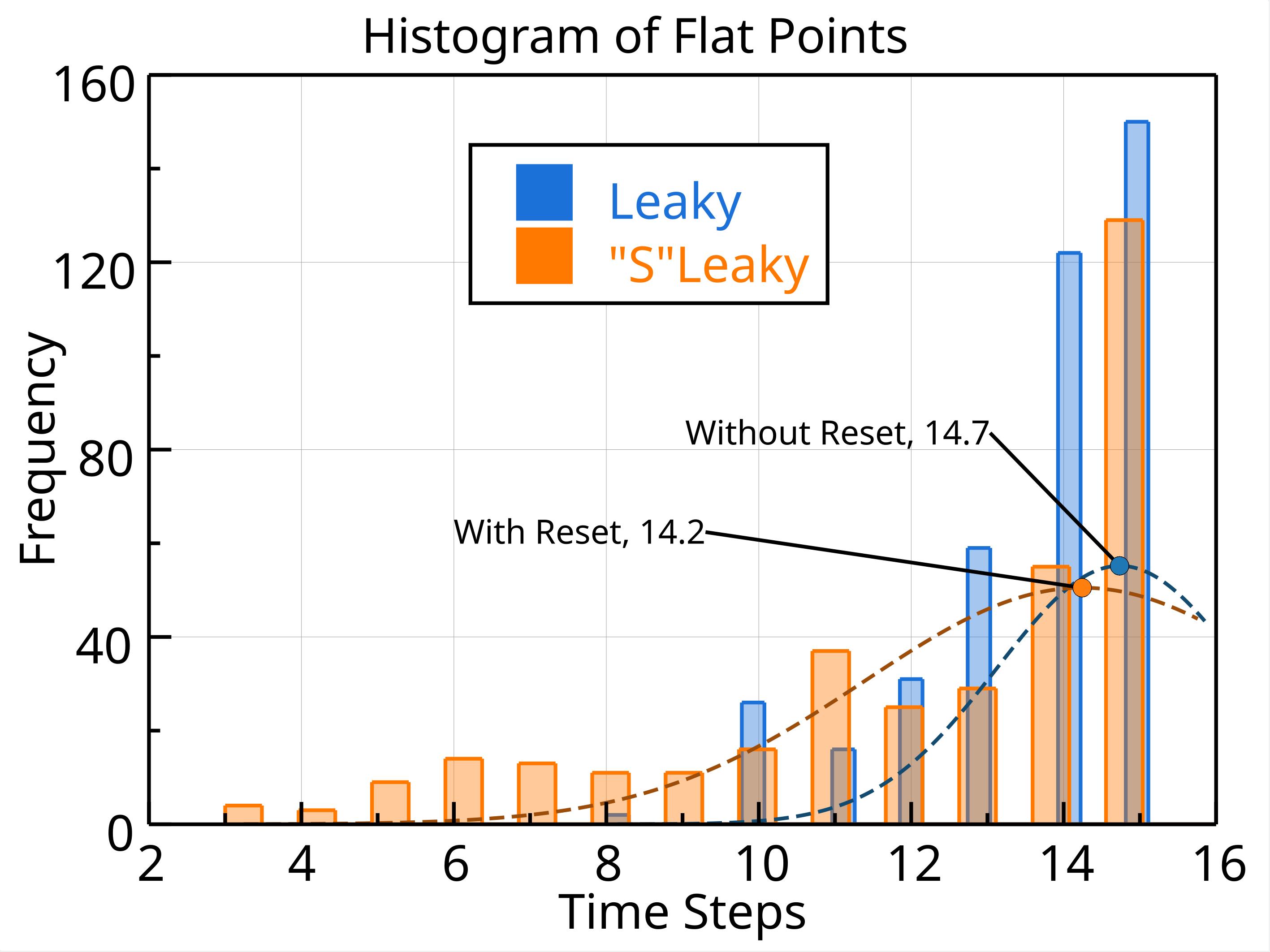}
    \caption{Distribution of stable points for Leaky versus. "S"Leaky neurons, showing earlier stabilisation with "S"Leaky units.}
    \label{fig:histogram}
\end{figure}

Since the stable points are not known until a complete forward pass is performed, we run inference on the network multiple times to assess the spread of stable points. Figure~\ref{fig:histogram} presents that data, comparing the distribution of stable points for an SNN with Leaky (with reset) and {\it SLeaky} (without reset) neurons. This provides evidence 
of how SPTTQ affects SNNs when a particular time step is chosen as the cut-off.

Based on the histogram, {\it SLeaky} on a standard SNN yields faster convergence of neurons to a stable state, allowing for an early SPTTQ cut-off. The mean stable point of SNN with Leaky and {\it SLeaky} neurons is 14.7 and 14.2 time steps, respectively. Although the difference is not significant, the {\it SLeaky}-based SNN had stable points within 10 steps, whereas the Leaky-SNN had nearly none. This is indicated by the variance of the normal distribution fit on histograms.

To evaluate this behaviour, we performed inference on the baseline ANN actor, SNN actor model from the HSAC algorithm, and SNNs with SPTTQ at six steps and {\it SLeaky} tweaks to compare the generated rewards. Table~\ref{tab:spikeSavins} presents the 50-episode average of rewards when the best actor model is used in the KENV environment.

Firstly, ANN and SNN models without any tweaks performed very closely. Next, when SPTTQ is applied at six steps on Leaky-SNN, the reward drops significantly. This is due to some neurons' inability to reach the required spike activity in time. From Figure~\ref{fig:histogram}, it was observed that Leaky-SNN had no stable points under 6 time steps, which explains this behaviour. But {\it SLeaky}-SNN performed nearly at the same level as an ANN; the relatively small number of stable points below 6 was sufficient to bring performance back to the same level.

Providing less time to finalise the output results in suboptimal control values. Despite skipping 10 steps, down from the initial 16, the nature of RL allowed the system to correct itself from the suboptimal action performed previously. As the number of skipped steps increases in SPTTQ, the number of mistakes increases, making the actor network exhibit a progressively worse action policy. This will be exhibited and discussed later.

\begin{figure}[ht!]
    \centering
    \includegraphics[width=1\linewidth]{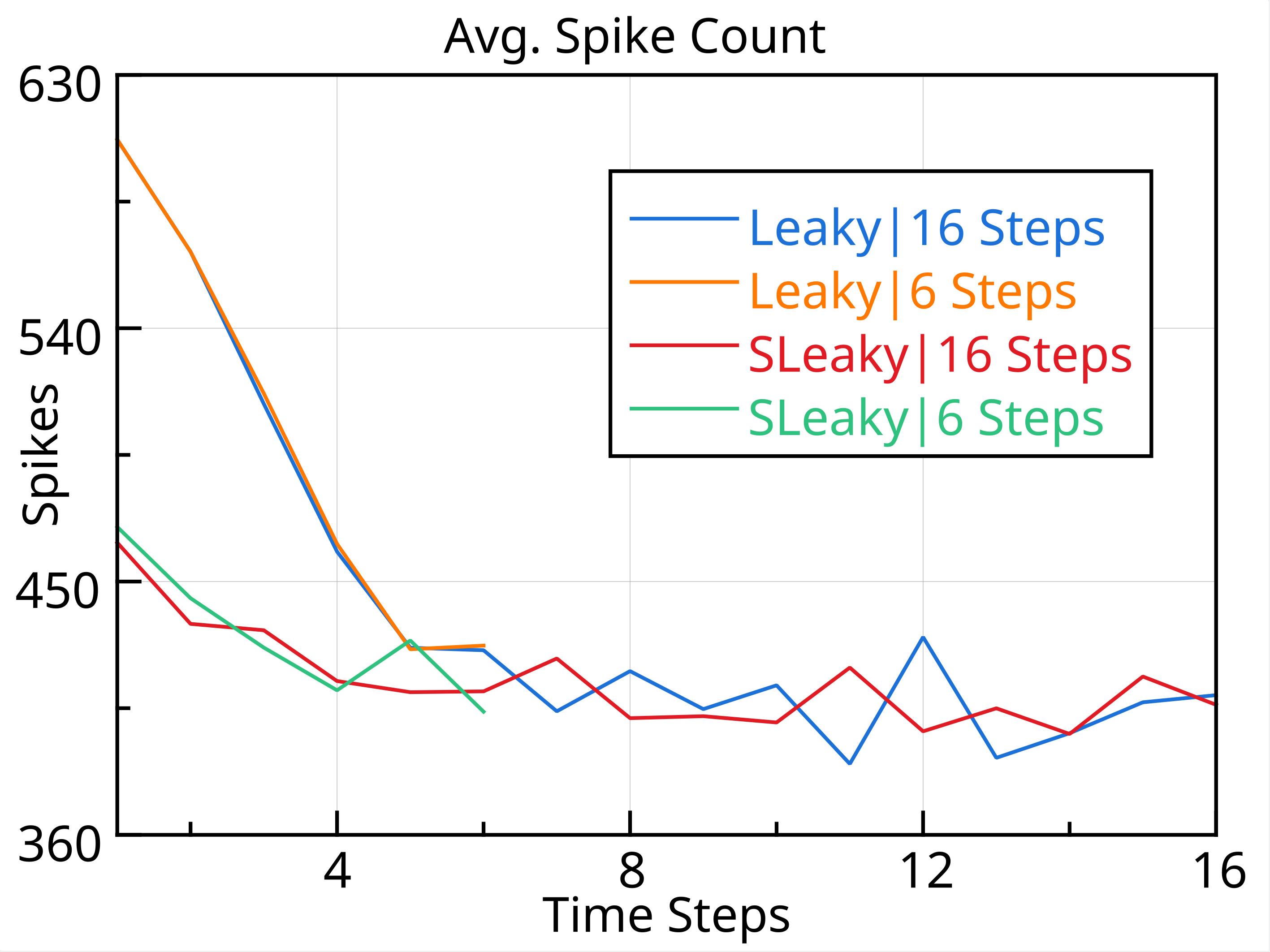}
    \caption{Average spike counts for different SNN variants, highlighting power reduction with {\it SLeaky} neurons.}
    \label{fig:avgSpikes}
\end{figure}

With 10 fewer time steps, each RL step takes 62.5\% less time, totalling a similar savings in overall inference latency. In terms of power, SNNs consume energy for each synaptic activity, i.e., when a spike is emitted by a neuron. To assess the power saved, we measured and compared the total number of spikes emitted by the network at each time step during one RL step. These numbers are averaged over multiple RL steps and KENV inference episodes. Figure~\ref{fig:avgSpikes} plots the average number of spikes emitted by the entire SNN at different time steps. All 4 combinations, with and without - {\it SLeaky} and SPTTQ, are compared in this chart.

From the plot, we observe that SNN models with Leaky neurons had high spiking activity in the initial time steps, reaching 600 spikes, but later lowered to 400. By swapping the neurons to {\it SLeaky}, the initial spiking activity also dropped to around 450, while later activity was retained as it is. The high initial spike activity was attributed to the SNN charging the deeper layer to the optimal spike rate. With {\it SLeaky}, all neurons are expected to contain some charge, hence SNNs need not spike as much and have to adjust minimally, leading to lower spike activity. Temporally, continuing with the SLeaky neuron places the neuron states at a later time step, which makes the network emit fewer spikes as seen in Figure~\ref{fig:avgSpikes}. The plot also includes SNNs with an SPTTQ cut-off of 6 steps, which exhibited spike-rate trends similar to those of their 16-step variants.


\begin{table*}[ht!]
\renewcommand{\arraystretch}{1.5}
    \centering
    \caption{Performance Comparison on KENV environment.}
    \label{tab:spikeSavins}
    \begin{tabular}[width=1\linewidth]{|c|c|c|c|c|c|c|c|c|}
        \hline
        \textbf{Model} & \textbf{Time Steps} & \textbf{Neuron} & \textbf{Total Spikes} & \textbf{Power Decrement\%} & \textbf{Total Time Steps} & \textbf{Latency Decrement\%} & \textbf{Reward} & \textbf{Trained} \\
        \hline
        ANN ASAC & - & ReLU & - & - & - & - & 3711 & $\checkmark$ \\
        \hline
        SNN SSAC~\cite{actorCriticSNN-sacSS-junqi} & 16 & Leaky & 1745 & - & 12000 & - & 2614 & $\times$ \\
        \hline
        SNN HSAC & 16 & Leaky & 7055 & Baseline & 12000 & Baseline & 3749 & $\checkmark$ \\
        \hline
        SNN HSAC & 16 & {\it SLeaky} & 6633 & 5.98\% & 12000 & 0 & 3685 & $\checkmark$ \\
        \hline
        SNN HSAC & 6 & Leaky & 3007 & 57.37\% & 4510 & 62.41\% & 2635 & $\times$ \\
        \hline
        SNN HSAC & 6 & {\it SLeaky} & 2584 & \textbf{63.37\%} & 4510 & 62.41\% & 3659 & $\checkmark$ \\
        \hline
    \end{tabular}
\end{table*}

Table~\ref{tab:spikeSavins} provides the total spikes emitted by each network configuration for one RL step. This is equivalent to the sum of all the points for each line in Figure~\ref{fig:avgSpikes}. First, the stock SNN configuration, with Leaky neurons and 16 time steps, amounted to over 7000 spikes. With the addition of {\it SLeaky}, the total dropped to 6633, representing 6\% of the power savings for similar performance. Although SNN with only SPTTQ cut-off of 6 steps exhibited 57\% spike savings, it performed poorly. Adding {\it SLeaky} not only improved its performance but also increased power savings to 63\% when compared with the original HSAC-trained SNN.


Actors controlling KENV will achieve a minimum reward value of 2600 due to the environment design, irrespective of how well the actor network is trained.
The control space design for the KENV actor network is constrained to move only forward, and the control values are also clipped into a safe range. Since the state space of KENV is discrete and linear, with 750 steps to mimic actual hardware, any random action control will result in the environment terminating successfully. Based on several tests, we found that an actor that is not even trained at all achieves a minimum reward of 2600. The performance continued to improve as the actor network was trained. This behaviour can be observed in Figure~\ref{fig:kenvTrain}. Hence, for this particular environment, we consider an actor that produces a final reward of around 2600 to be untrained, as marked under the {\it Trained} column in Table~\ref{tab:spikeSavins}.


\begin{figure}[ht!]
    \centering
    \includegraphics[width=1\linewidth]{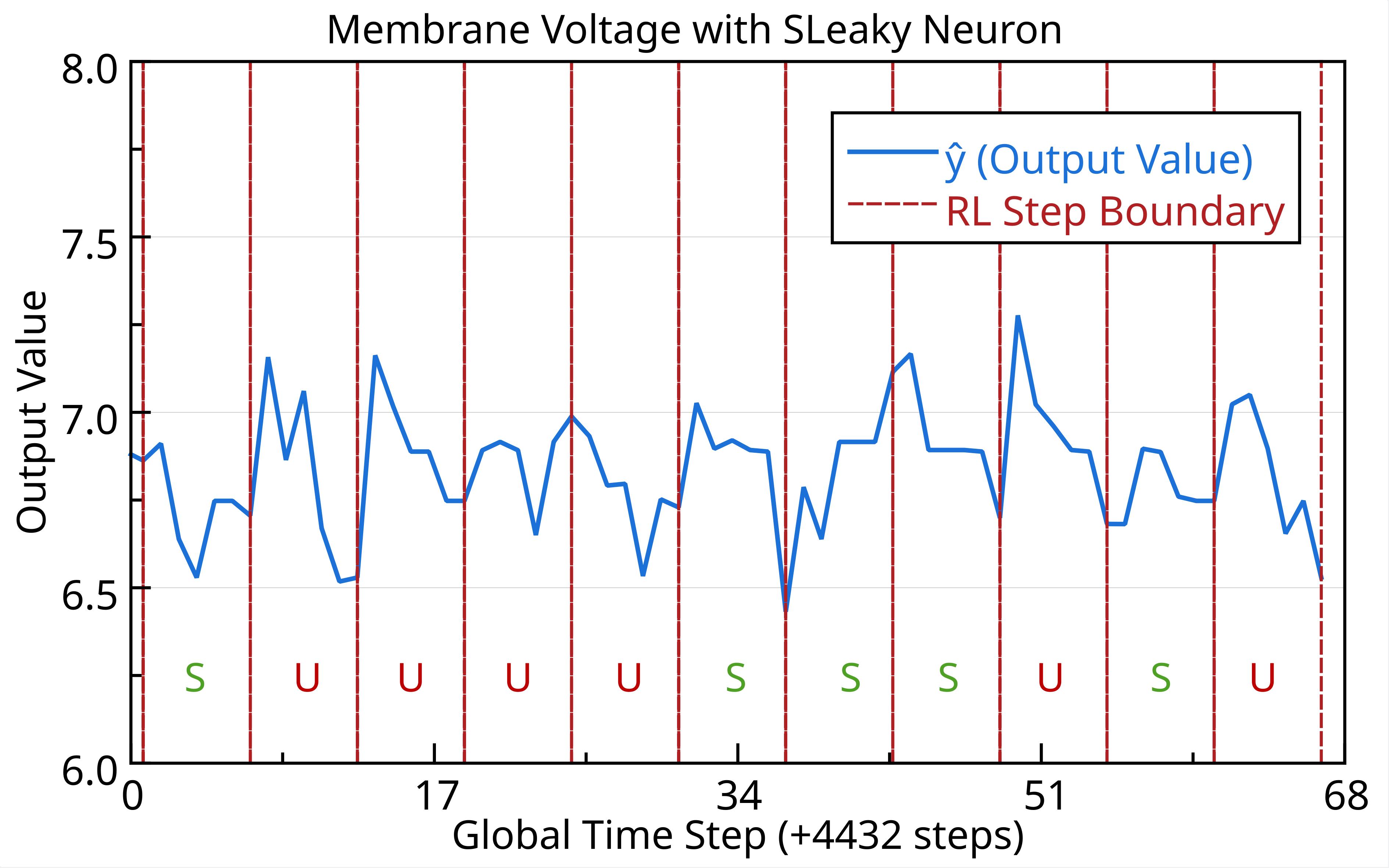}
    \caption{Output evolution across time steps under SPTTQ, with stable and unstable states marked for each RL step.}
    \label{fig:flowOfMembrane}
\end{figure}

Figure~\ref{fig:flowOfMembrane} shows how the output values of an SNN with SPTTQ at six steps and {\it SLeaky} neuron changes at each time step. The vertical red lines represent RL step boundaries, with a 6-time-step gap between them. Stable and unstable states are marked using a green S and red U, respectively.

\subsection{Post Training Tweaks}
\begin{figure*}[htb!]
    \centering
    \includegraphics[width=1\linewidth]{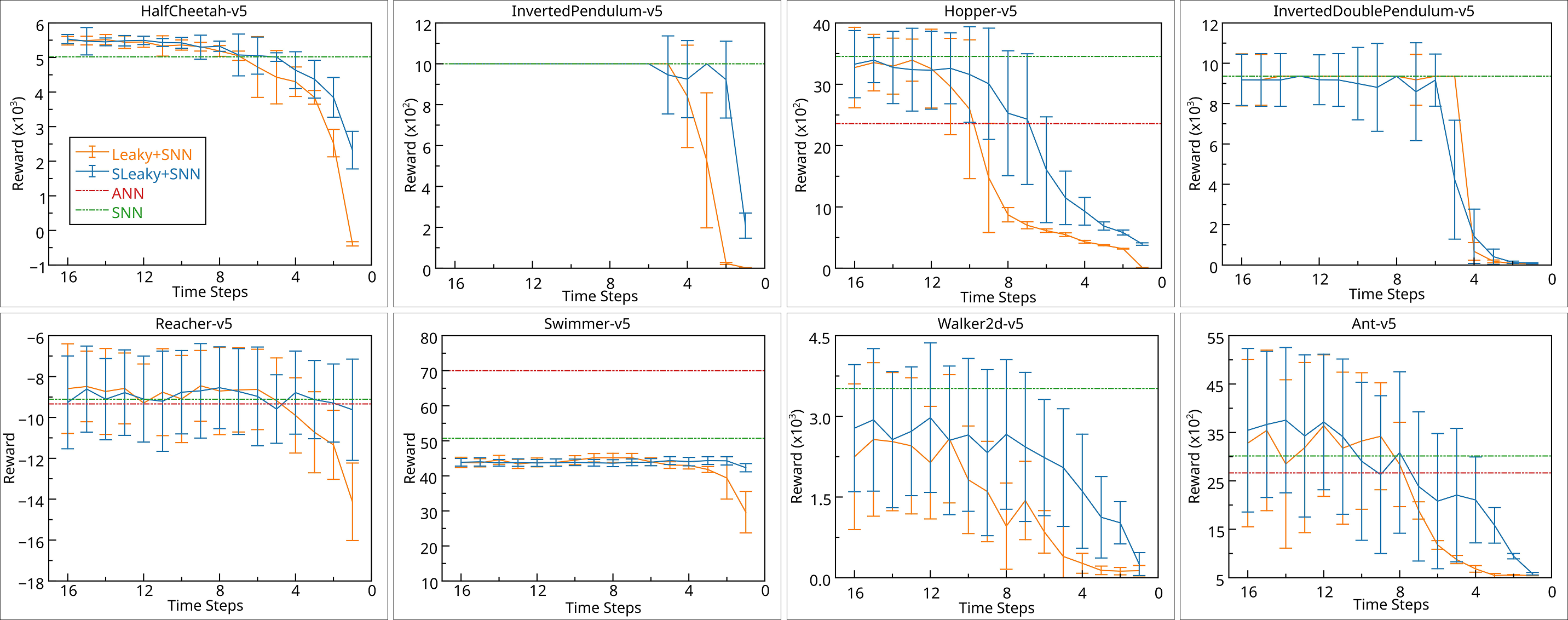}
    \caption{Average episodic rewards across different SPTTQ cut-off steps. Performance drops as fewer internal steps are used, with simpler environments showing sharper declines, and {\it SLeaky} generally performs more robustly than Leaky.}
    \label{fig:infRew}
\end{figure*}

\begin{figure*}[ht!]
    \centering
    \includegraphics[width=1\linewidth]{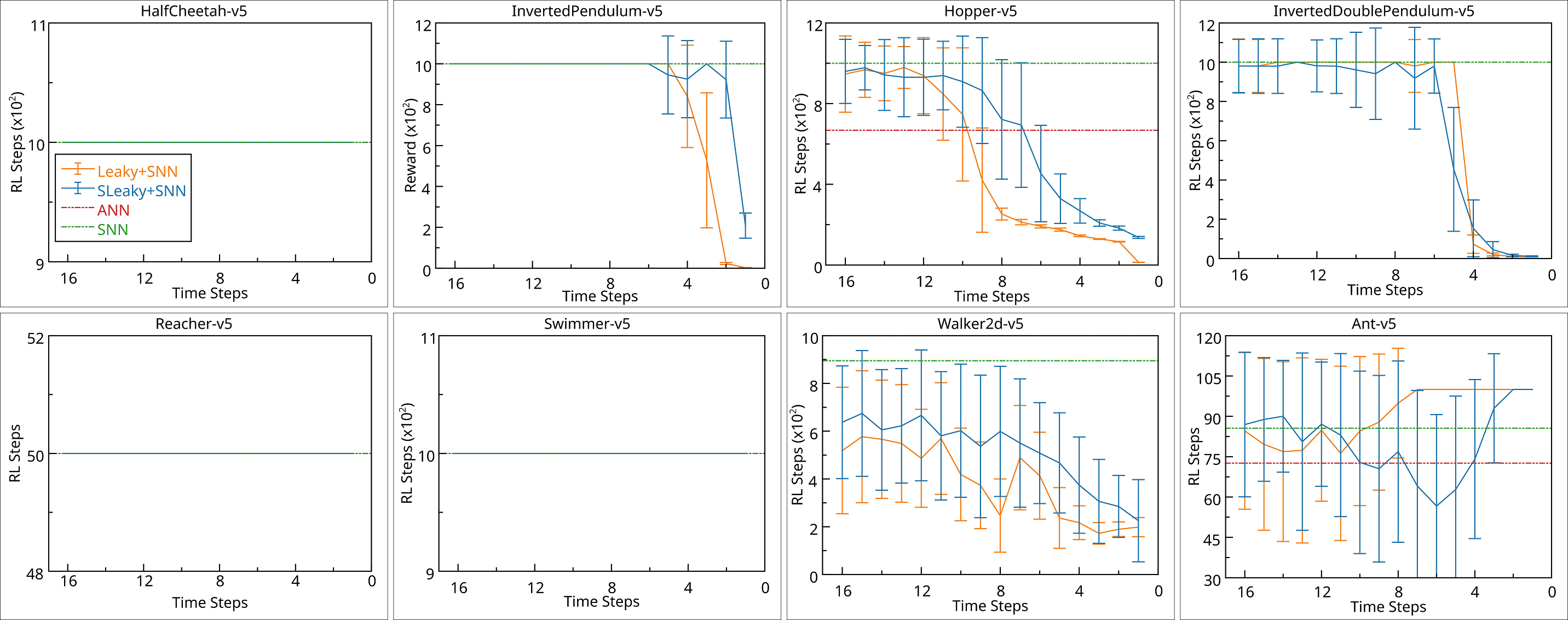}
    \caption{Average RL steps for each cut-off setting. Trends mostly mirror the reward curves.}
    \label{fig:infStep}
\end{figure*}
Based on the behavioural and performance impacts of SPTTQ and {\it SLeaky} on a trained SNN, we analyse their progressive performance impact on multiple environments, starting with the same MuJoCo environments used for HSAC. For this, SNN actor models trained with HSAC are subjected to the optimisation flow in Algorithm~\ref{alg:optSNN}. They are subjected to inference \& record the total episodic return and runtime. As a comparison, we also include the results of the ANN actor from ASAC and the unmodified SNN from HSAC.

Figures~\ref{fig:infRew} and \ref{fig:infStep} present total episodic returns and time steps, averaged over 50 episodes. Time steps are calculated until termination or truncation. The x-axis represents the SPTTQ time step, and vertical lines represent the episodic inference variance of reward and RL steps. Overall, these graphs show a drop in performance as the SPTTQ cut-off is varied from $T$ to 1.

From the reward graphs, we observe that performance progressively drops as the time-step cut-off decreases. Simple environments showed a distinct drop in reward after a certain cut-off, whereas the others exhibited a gradual reduction in performance. Between the neurons, {\it SLeaky} showed similar or better performance and resilience towards SPTTQ than Leaky. While most environments performed well at high cut-off steps, some showed anomalous behaviour. Firstly, {\it InvertedDoublePendulum} exhibited greater variance and, occasionally, worse performance with {\it SLeaky} neurons compared to Leaky neurons. Next, the Swimmer environment, despite exhibiting higher reward values, failed to learn the task. This explains the wide gap between the ANN reward line at 70 and all SNNs, which are below 50.
Other than some graphs, the RL step plots also present a similar trend. Of the anomalies, {\it HalfCheetah}, {\it Reacher}, and {\it Swimmer} do not support termination conditions, and hence always hit their truncation limit. Due to the simple reward design, the reward and step graphs of {\it InvertedPendulum}, {\it InvertedDoublePendulum}, {\it Hopper}, and {\it Walker2d} coincide.

\begin{figure}[ht!]
    \centering
    \includegraphics[width=1\linewidth]{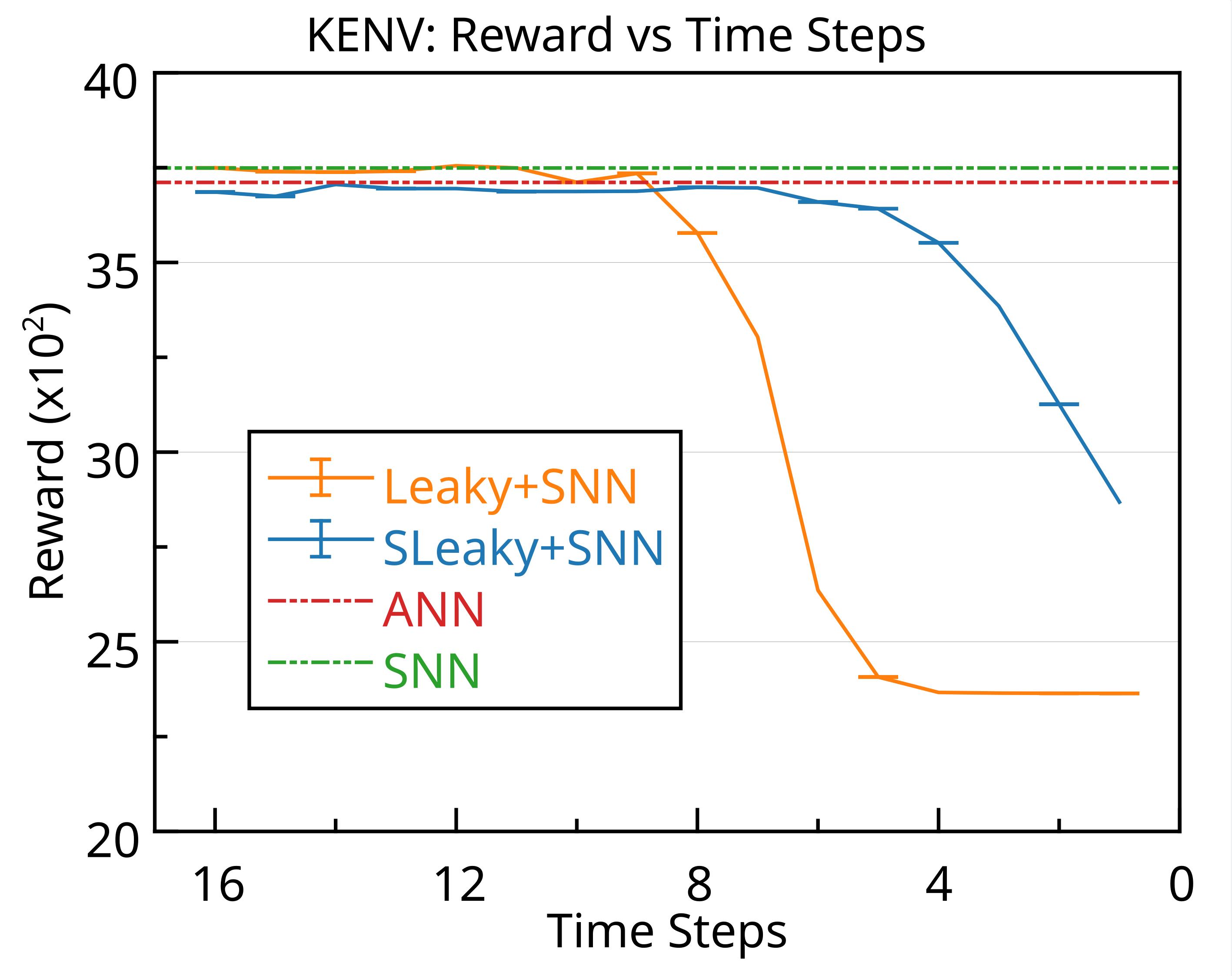}
    \caption{KENV rewards under different SPTTQ cut-off steps. {\it SLeaky}-SNN stays close to ANN and degrades more gradually, while Leaky-SNN falls off earlier.}
    \label{fig:kenvInfRew}
\end{figure}

\begin{figure}[ht!]
    \centering
    \includegraphics[width=1\linewidth]{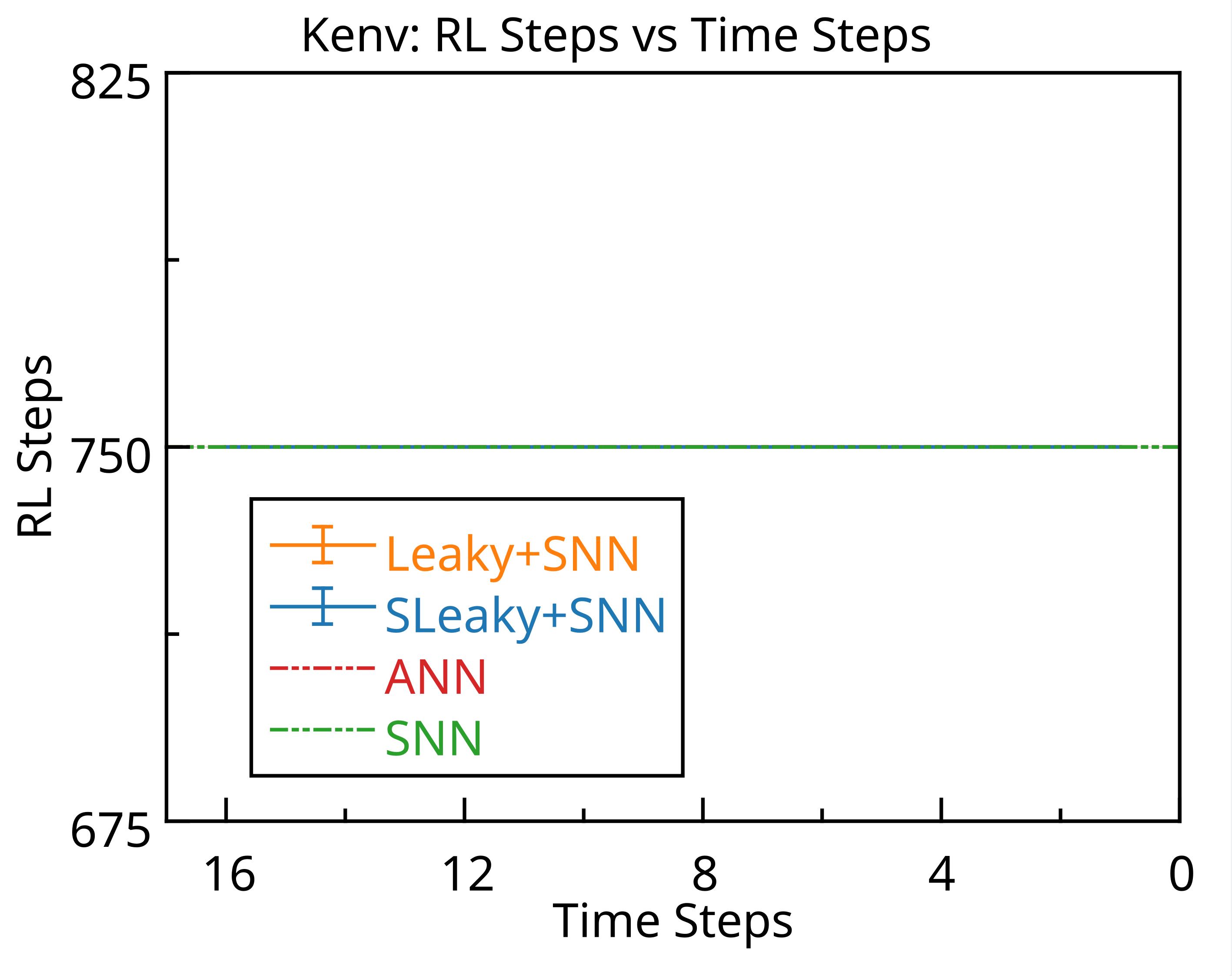}
    \caption{RL steps in KENV across cut-off steps. Since KENV has no terminating condition, all agents run the full 750 steps regardless of neuron type or cut-off.}
    \label{fig:kenvInfStep}
\end{figure}

While the number of steps and reward decreased as the cut-off step decreased,
the Ant environment's RL steps increased after a certain point. This happens since the termination conditions require the Ant model to tilt or flip over. The reward in this environment comprises a forward reward, an alive reward, and a control cost. Since a bad action policy was unable to move or terminate itself, the agent mostly remained stationary, accumulating steps but not enough positive reward. As policy worsened, reward dropped predictably, but the total number of steps in this particular environment increased.

Applying the same methodology to KENV, we get results as shown in Figures~\ref{fig:kenvInfRew} and \ref{fig:kenvInfStep}. {\it SLeaky}-SNN outperformed Leaky-based SNN, with the highest gains at SPTTQ between 7 and 4 steps. Since KENV also does not support a termination state, the environment always runs for 750 RL steps.

\begin{figure}[ht!]
    \centering
    \includegraphics[width=1\linewidth]{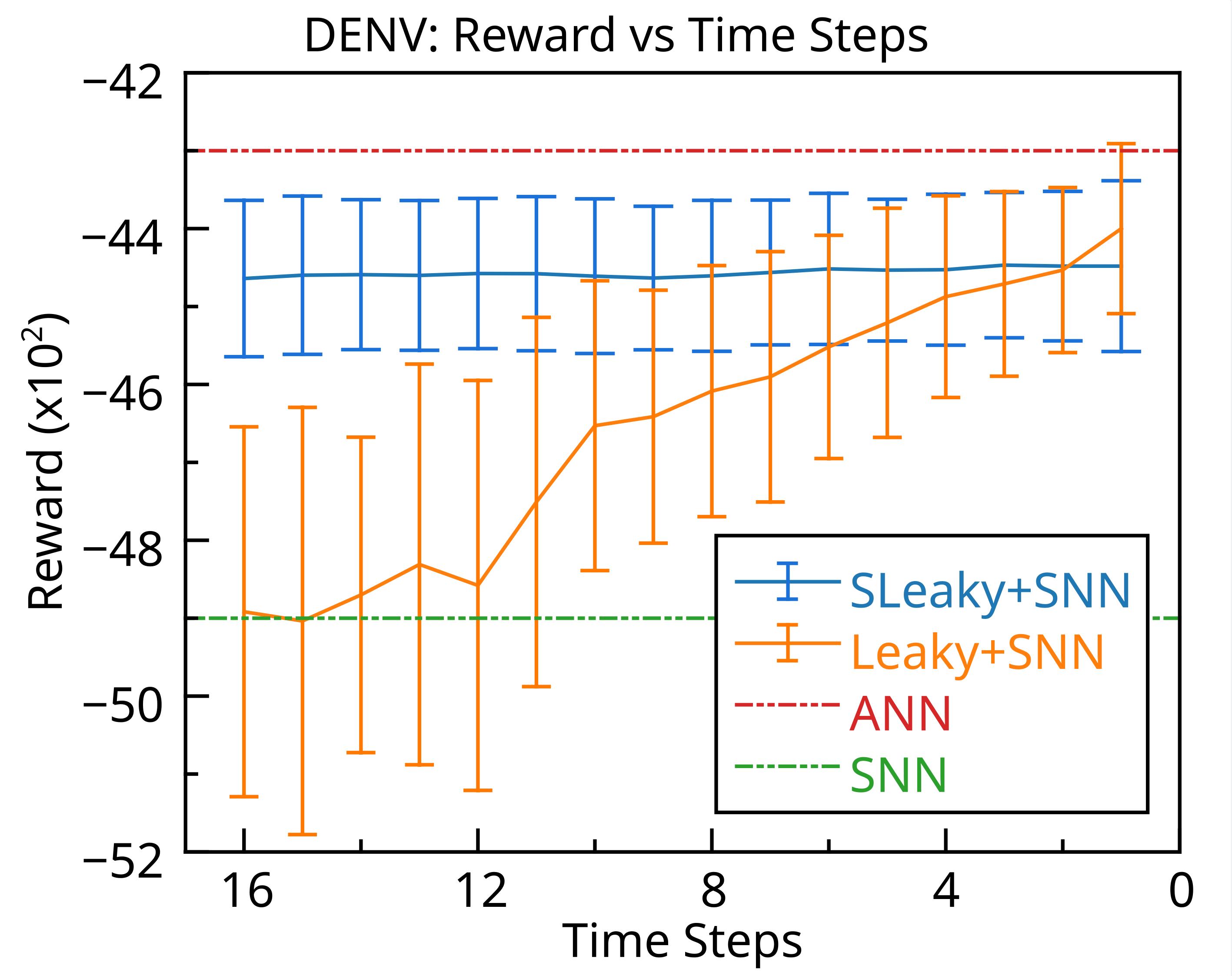}
    \caption{DENV rewards for ANN, Leaky-SNN, and {\it SLeaky}-SNN at different cut-off settings. {\it SLeaky} tracks ANN most closely, while Leaky shows greater variation and decline.}
    \label{fig:denvInfRew}
\end{figure}

\begin{figure}[ht!]
    \centering
    \includegraphics[width=1\linewidth]{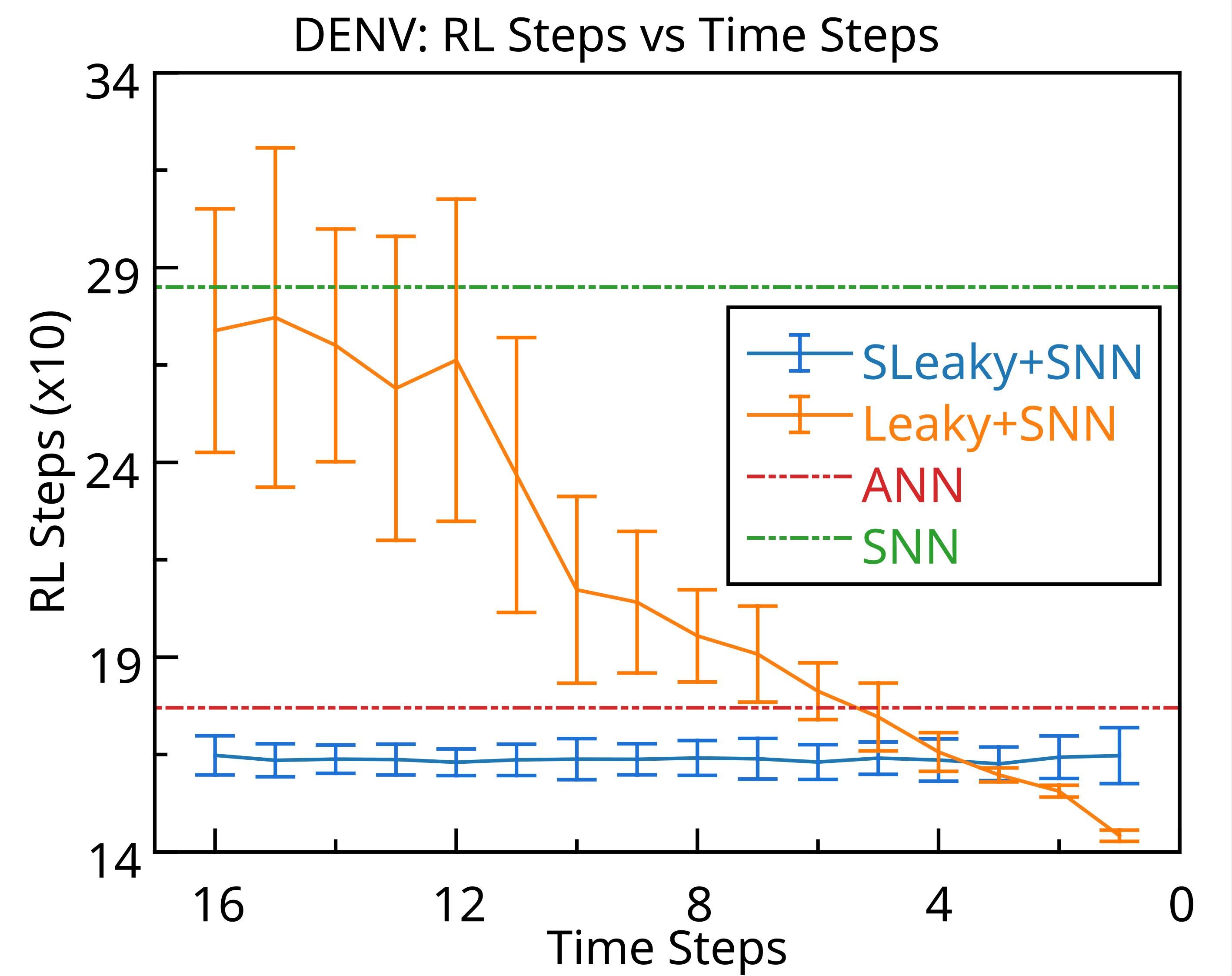}
    \caption{DENV RL steps across cut-off settings. As the cut-off decreases and the policy weakens, Leaky-SNN tends to take longer to finish episodes, while {\it SLeaky} remains closer to ANN behaviour.}
    \label{fig:denvInfStep}
\end{figure}

Likewise, Figures~\ref{fig:denvInfRew} and \ref{fig:denvInfStep} represent inference performance of DENV. This environment exhibited a stark difference from the base SNN and ANN behaviours, despite having a smooth action policy. The ANN-based actor network completed an iteration within 180 RL steps, whereas the SNN actor took nearly 290 RL steps. Most of the additional steps taken by the SNN actor were towards the end of the episode, where the arm structure moved more slowly. SNN with Leaky behaved like a normal SNN with all 16 steps, but steadily drifted towards ANN behaviour as the SPTTQ cut-off decreased to lower steps. On the other hand, the Sequent Leaky variant consistently performed on par with the original ANN, as evident by the reward and step graphs.

\subsection{Simulation and Control Behaviour}

This subsection visually presents the environments' dynamics by plotting simulation parameters during inference. This explains the effect of the reward components 
on the control behaviour.

\subsubsection{Dynamic Environment}

\begin{figure}[ht!]
\centering
\includegraphics[width=1\linewidth]{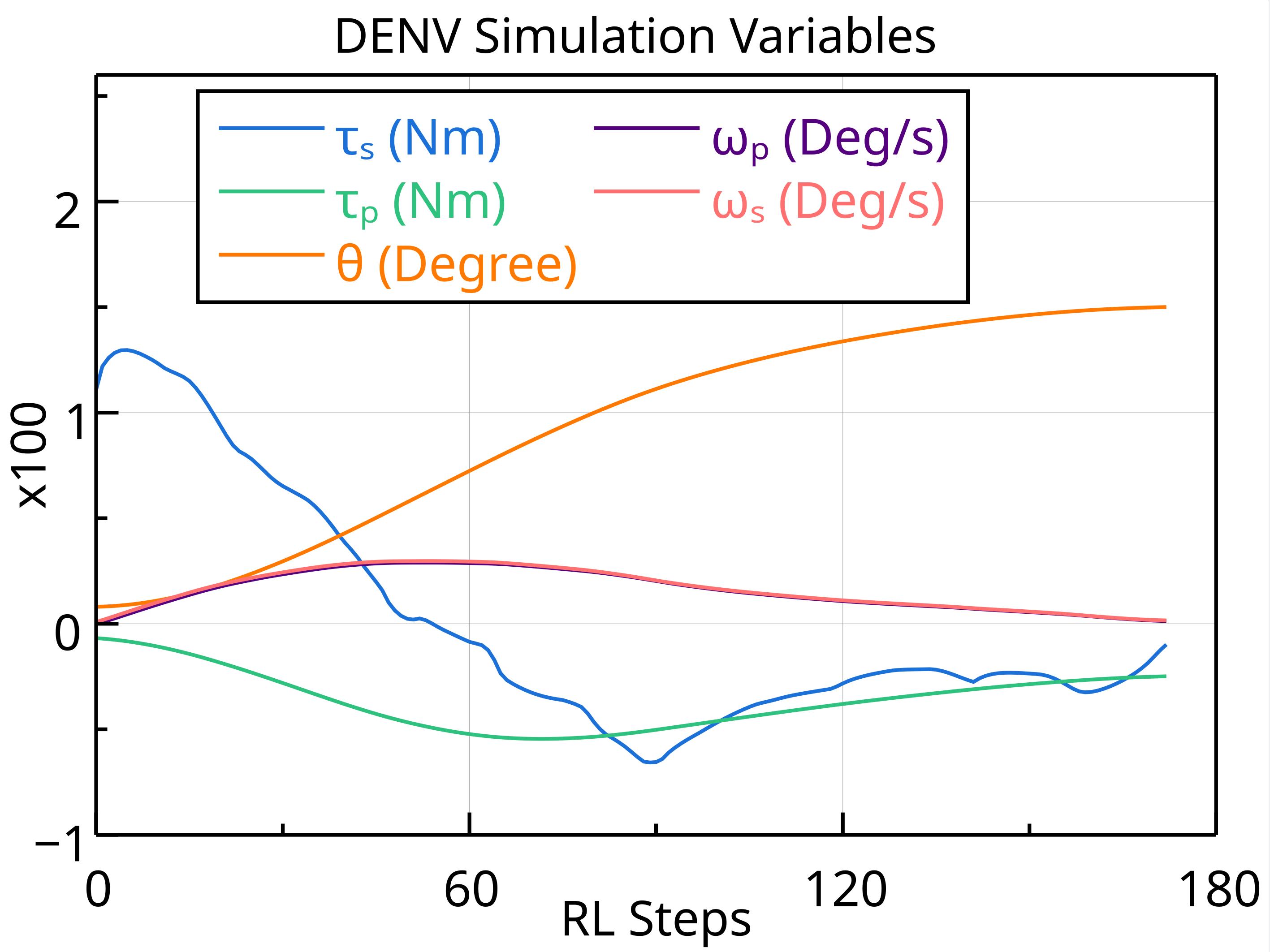}
\caption{Simulation variables of the dynamic environment. Y-axis is unit agnostic; units for each line are provided in the plot's legend.}
\label{fig:denvSimComps}
\end{figure}

\begin{figure}[ht!]
\centering
\includegraphics[width=1\linewidth]{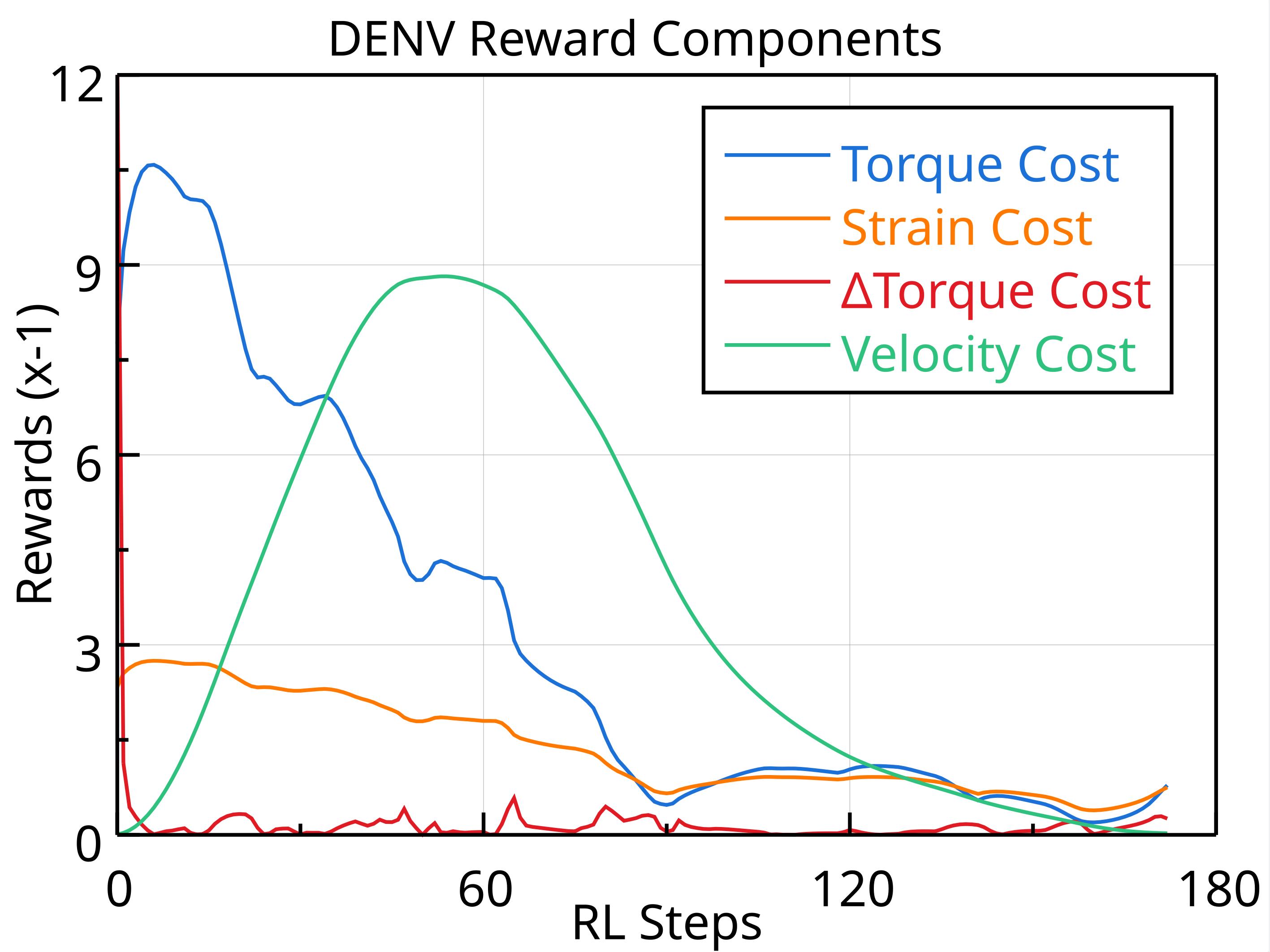}
\caption{Reward components during inference on DENV.}
\label{fig:denvRewComps}
\end{figure}

DENV predicts the system's target torque for a given time step, which is then translated into hardware steps and delays. Figure~\ref{fig:denvSimComps} plots those computed torque components to simulate the system. Strain on the patient is derived from the difference between the hand velocity $\omega_p$ and the system velocity $\omega_s$. Essential observations include the smoothness of system torque $\tau_s$ enforced by its reward component as shown in Figure~\ref{fig:denvRewComps}, and minimisation of velocity at episode termination.

\subsubsection{Kinematic Environment}

KENV predicts the delay pattern between stepper motor commands at each RL step, which in turn determines the angular progression of the arm.

\begin{figure}[ht!]
\centering
\includegraphics[width=1\linewidth]{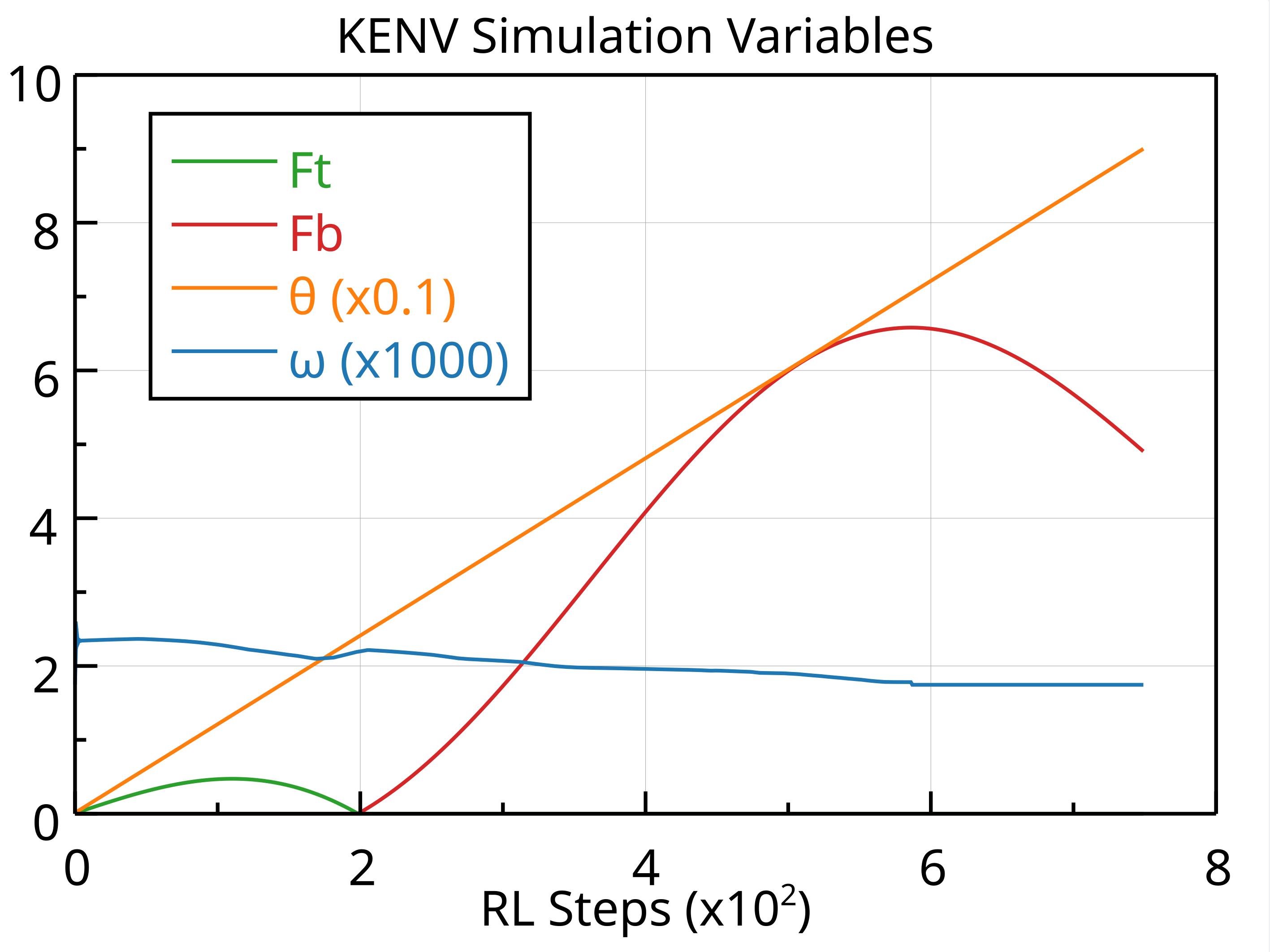}
\caption{Simulation variables of the KENV environment. The legend contains the unit of each variable.}
\label{fig:kenvSimComps}
\end{figure}

\begin{figure}[ht!]
\centering
\includegraphics[width=1\linewidth]{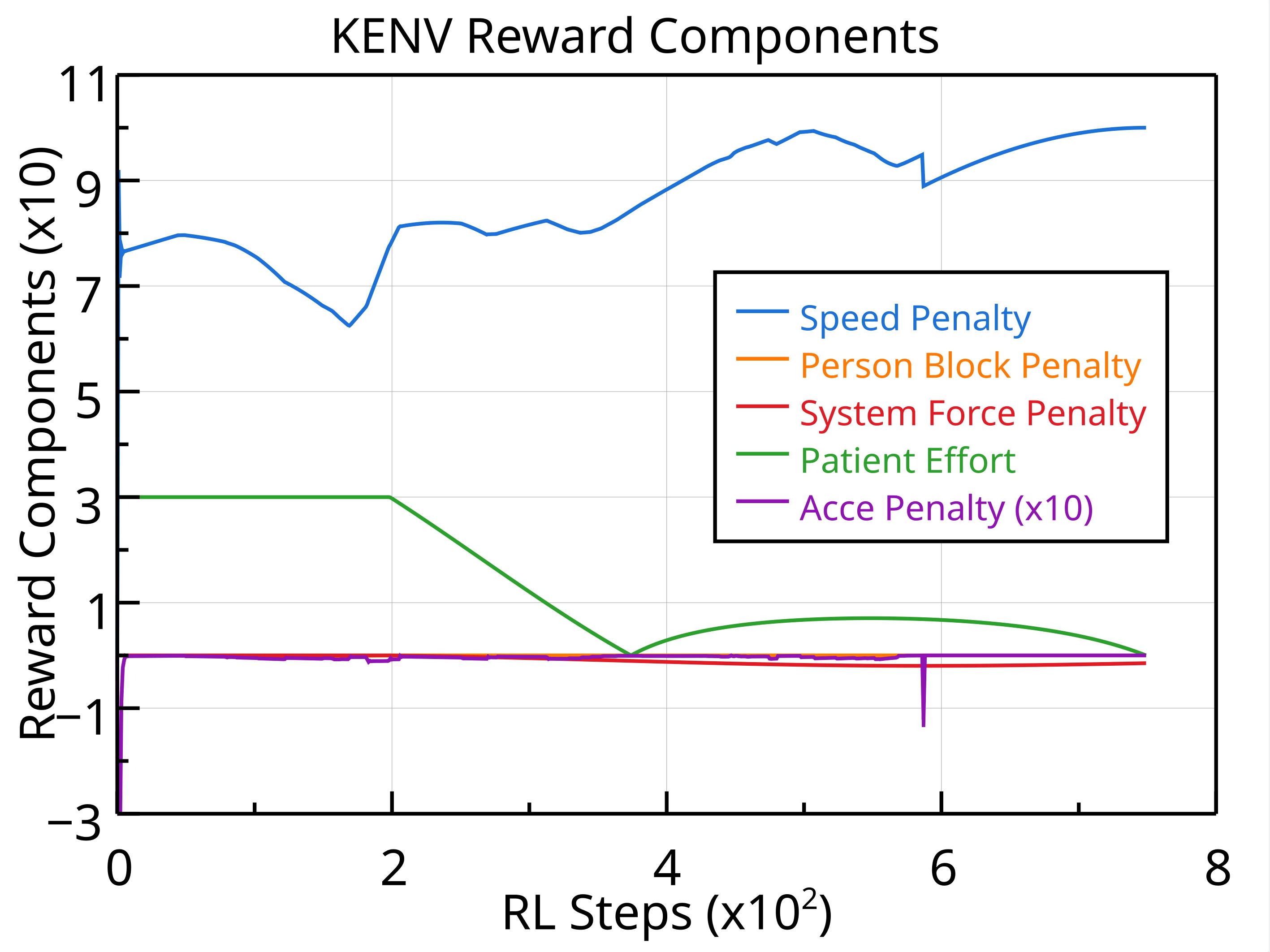}
\caption{Reward components of KENV during inference.}
\label{fig:kenvRewComps}
\end{figure}

Figures~\ref{fig:kenvSimComps} and~\ref{fig:kenvRewComps} show the trained policy behaviour in KENV. In Figure~\ref{fig:kenvSimComps}, the joint angle $\theta$ increases smoothly over time while the angular velocity $\omega$ gradually decreases, so the arm speeds up early and slows near the end of the motion. The total force on the patient $F_t$ stays low, and the system force $F_b$ follows a single rise and fall, indicating monotonic, well-controlled movement without overshoot.

Figure~\ref{fig:kenvRewComps} decomposes the reward. The speed term plays a dominant role, indicating the policy’s accuracy in tracking the target speed profile. Patient effort starts high and then reduces as the system takes over, while the block and force penalties remain near zero, which means that the controller rarely pushes against the patient or applies excessive force. The acceleration penalty only shows a brief spike, confirming that delay changes are mostly smooth.

\sectionEND{Experiments, Results and Discussion}


\section{Conclusion}
This work presented \textit{NeuRehab}, an end-to-end framework that combines reinforcement learning and spiking neural networks to enable autonomous, resource-aware rehabilitation on a wheelchair-mounted exoskeleton. We grounded our design on the XoRehab platform and introduced two complementary simulation environments, KENV and DENV, that capture the discrete, stepper-motor kinematics and continuous torque-driven dynamics of the shoulder joint. These environments allowed us to safely train and analyse deep RL policies before deployment, while explicitly encoding clinically motivated constraints such as strain minimisation, smooth motion, and assistance-as-needed behaviour.

On the algorithmic side, 
we proposed the Hybrid-SAC (HSAC) scheme, which uses a spiking actor and an ANN critic to align training with the eventual neuromorphic deployment. Across standard MuJoCo benchmarks as well as KENV and DENV, HSAC reliably tracked the performance of a full-precision ANN SAC (ASAC) and outperformed an all-spiking SAC (SSAC), demonstrating that a heterogeneous actor–critic split can provide a favourable balance between precision, sample efficiency, and hardware compatibility.

We further introduced two inference-time optimizations tailored to continuous control: Spiking Post-Training Temporal Quantisation (SPTTQ) and the Sequent Leaky ({\it SLeaky}) neuron. SPTTQ treats the number of spiking time steps as a post-training quantisation knob, selecting an early cut-off that preserves control performance, while {\it SLeaky} reuses membrane states across RL steps to reduce charge-up overhead. Together, these techniques reduced spike counts by up to $\sim$63\% and improved effective latency by more than 60\% on KENV, while maintaining rewards close to the ANN baseline, and showed improved robustness to temporal truncation across several MuJoCo tasks.

Overall, our results indicate that neuromorphic reinforcement learning for rehabilitation is both feasible and practical when the algorithm, simulator, and hardware are co-designed.

\sectionEND{Conclusion}

\section{Future Work}

As a modular framework, the proposed system can be applied to other actuation joints in the XoRehab system. Since all joints operate independently, NeuRehab can be extended to other joints by adding their mathematical simulation to the pipeline. NeuRehab can also work with any hardware configuration provided that it satisfies key design constraints, such as the processing architectures.

The test parameters can be further broadened to include more human torque profiles. This adds one more input variable for the actor to learn. Based on limited testing, we found that the actors learned only with transfer learning from base actor networks. Hence, the system can contain a basis control profile, which learns the patient's behaviour.

SNNs and OWS decoder enable the software's applicability to other time and power-constrained environments, such as industrial machines. More neurons, such as the Current-Based Leaky or Adaptive Leaky neurons, can be investigated further to process data with enhanced sparsity. While this specific work doesn't benefit from rate encoding due SPTTQ, early exit when a neuron's spike rate stabilises can be tested.

\sectionEND{Future Work}

\appendices 

\section{Abbreviations and Common Terms}

\begin{itemize}
    \item \textbf{RL}: Reinforcement learning
    \item \textbf{SAC}: Soft actor critic RL algorithm
    \item \textbf{ANN, SNN}: Artificial neural network and Spiking neural network.
    \item \textbf{Leaky}: Refers to the first order Leaky-Integrate-and-Spike (LIF) neuron, can be used interchangeably
    \item \textbf{RL steps}: Refers to the individual RL time steps of being in a state.
    \item \textbf{Time steps}: Refers to the individual spiking time steps discretised by spike encoding.
    \item \textbf{Forward pass}: The process to calculate the final predicted value for a given input value at a time step. In ANNs, this runs the network computation once; in SNNs, it runs $T$ times, once for each encoded time step from an RL step.
    \item \textbf{Arm Structure}: Refers to the arm hardware fixture, also known as the arm part and the arm system.
    \item \textbf{Input}: Refers to the input to a forward pass process, which are observations from an environment. Represents the non-encoded value in the spiking domain.
    \item \textbf{Membrane potential}: Represents the membrane voltage, used interchangeably in the manuscript.
    \item \textbf{ASAC, HSAC and SSAC}: 3 modes of training: ASAC is the all ANN variant, SSAC is the baseline with all spiking networks, and HSAC is the proposed structure with one SNN actor and one ANN critic.
    \item \textbf{KENV and DENV}: Abbreviated form of Kinematic and Dynamic Environments.
\end{itemize}

\section{Hyperparameters}
\label{app:netShapes}
All neural networks use multilayer perceptrons with ReLU activations and shared hidden sizes across the actor and critic.

\subsection{Actor Network}
\label{app:actor}
The actor network outputs the parameters of a squashed Gaussian policy. The architecture is:
\begin{itemize}
    \item Input: state vector $s_t$
    \item Layer 1: fully connected, 512 units, ReLU
    \item Layer 2: fully connected, 512 units, ReLU
    \item Layer 3: fully connected, 384 units, ReLU
    \item Output: mean and log standard deviation for each action dimension
\end{itemize}
The sampled action is passed through a $\tanh$ nonlinearity and then rescaled to the environment action bounds using the affine transformation stored in the \texttt{action\_scale} and \texttt{action\_bias} buffers. The spiking variants use Leaky or {\it SLeaky} spiking neurons in place of ReLU, and 2 OWS decoders for the output layers for action mean and variance.

\subsection{Critic Networks}
\label{app:critic}
SAC uses two independent Q-networks $Q_{\theta_1}(s,a)$ and $Q_{\theta_2}(s,a)$. Each critic has the following structure:
\begin{itemize}
    \item Input: concatenated state and action vector $(s_t, a_t)$
    \item Layer 1: fully connected, 512 units, ReLU
    \item Layer 2: fully connected, 512 units, ReLU
    \item Layer 3: fully connected, 384 units, ReLU
    \item Output: scalar Q-value
\end{itemize}
Target Q-networks have the same architecture and are updated by Polyak averaging~\cite{polyakAvg-polyakAvg-polyak} from the online critics. Similar to the actor network, spiking networks use Leaky and {\it SLeaky} instead of ReLU.

\section{Training Hyperparameters}
\label{app:hparams}
Table~\ref{tab:hyperparams_sac} lists the hyperparameters used by the SAC implementation in our experiments. These values follow the CleanRL SAC \cite{huang2022cleanrl} configuration, with minor adjustments for our custom environment.

\begin{table}[H]
\centering
\begin{tabular}{l c}
\hline
\textbf{Parameter} & \textbf{Value} \\ \hline
Total timesteps & 500{,}000 \\
Replay buffer size & 50{,}000 transitions \\
Discount factor $\gamma$ & 0.99 \\
Soft update coefficient $\tau$ & 0.005 \\
Batch size & 256 \\
Learning starts & 5{,}000 steps \\
Actor learning rate (policy\_lr) & $3\times 10^{-4}$ \\
Critic learning rate (q\_lr) & $1\times 10^{-4}$ \\
Optimizer & Adam (all networks) \\
Policy update frequency & every 2 critic updates \\
Target network update frequency & every 3 steps \\
Initial entropy coefficient $\alpha$ & 0.2 \\
Entropy tuning & automatic (log $\alpha$ learned) \\
Action sampling & Gaussian, squashed by $\tanh$ \\
Loss function & Mean Squared Error \\
\hline
\end{tabular}
\caption{SAC hyperparameters used in the robot arm experiments.}
\label{tab:hyperparams_sac}
\end{table}

Table~\ref{tab:hyperparams_snn} lists the hyperparameters used for training and inference of spiking neural networks.
\begin{table}[H]
    \centering
    \begin{tabular}{l c}
        \hline
        \textbf{Parameter} & \textbf{Value} \\
        \hline
        Neurons & Leaky, {\it SLeaky} \\
        Reset mechanism & Zero-reset \\
        Total time steps & 16 \\
        Short time steps & varies \\
        Initial voltage decay $\beta$ & 1.0 \\
        Initial threshold $V_{th}$ & 2.0 \\
        Surrogate gradient function & Fast Sigmoid \\
        Surrogate slope & 10.0 \\
        Decay $\beta$ & Trainable \\
        Threshold $V_{th}$ & Trainable \\
        \hline
    \end{tabular}
    \caption{Spiking neural network hyperparameters used in spiking actors.}
    \label{tab:hyperparams_snn}
\end{table}

\section{Software Environment}

All the experiments are performed using Python 3.12, PyTorch 2.7, SNNTorch 0.7, Gymnasium 1.2, and Stable Baselines 3. The training hardware consists of an Nvidia A4500 GPU and an AMD Ryzen 9 5950X CPU with 64GB of RAM running Fedora Linux 42.








\section*{Acknowledgment}

The authors would like to thank IIIT-Bangalore for facilitating this research. We have not received any funding specifically for this work. Special thanks to Prof. Raghuram Bharadwaj for assistance with reinforcement learning and for proofreading.

\bibliographystyle{ieeetr} 
\bibliography{citations.bib}




\balance

\end{document}